\documentclass{article}

\usepackage{amsmath}
\usepackage{amssymb}
\usepackage{theorem}
\usepackage{xspace}
\usepackage[hyphens]{url}
\usepackage{datetime}
\usepackage{authblk}
\usepackage[usenames,dvipsnames,svgnames]{xcolor}

\usepackage{algorithmic, algorithm}
\usepackage{cite}
\usepackage[normal]{subfigure}
\usepackage{siunitx}
\usepackage{microtype}
\usepackage[margin=1.8cm]{geometry}

\usepackage{epstopdf}

\usepackage{multirow}
\usepackage{slashbox}

\usepackage{tikz}
\usetikzlibrary{shapes,arrows}
\usetikzlibrary{decorations,calc}

\newcommand{\pinf}{\ensuremath{{+\infty}}}

\newcommand{\hFilterSet}{\ensuremath{\mathbb{H}}}

\newcommand{\RR}{\ensuremath{\mathbb{R}}}
\newcommand{\KK}{\ensuremath{\mathbb{K}}}
\newcommand{\RPP}{\ensuremath{\left]0,+\infty\right[}}
\newcommand{\RX}{\ensuremath{\left]-\infty,+\infty\right]}}

\newcommand{\bR}{\ensuremath{\mathbf{R}}}
\newcommand{\bh}{\ensuremath{\mathbf{h}}}

\newcommand{\I}{\ensuremath{\operatorname{I}\xspace}}

\newcommand{\db}[1]{\SI{#1}{\decibel}}

\newcommand{\abbFrame}{\ensuremath{f}} 
\newcommand{\abbBasis}{\ensuremath{b}}
\newcommand{\abbFrameVsBasis}{\ensuremath{\abbFrame{}/\abbBasis{}}} 

\newcommand{\tabEmphA}[1]{{\bfseries{#1}}} %
\newcommand{\tabEmphB}[1]{\textit{#1}} %
\newcommand{\tabEmphC}[1]{\textit{#1}} %

\def\ns#1{|\mskip -2 mu|\mskip -2 mu|#1|\mskip -2mu |\mskip -2mu|}

\graphicspath{{./Figures_color/}{./Figures_BW/}}

\title{A Primal-Dual Proximal Algorithm for Sparse Template-Based Adaptive Filtering: Application to Seismic Multiple Removal}
\author[1,3]{Mai Quyen Pham\thanks{mai-quyen.pham@ifpen.fr}}
\author[2]{Caroline Chaux\thanks{caroline.chaux@latp.univ-mrs.fr}}
\author[1]{Laurent Duval\thanks{laurent.duval@ifpen.fr}}
\author[3]{Jean-Christophe Pesquet\thanks{pesquet@univ-mlv.fr}}
\affil[1]{IFP Energies nouvelles, Dir. Technologie\\ 
1 et 4 Av de Bois-Préau, 92852 Rueil-Malmaison Cedex - France}
\affil[2]{Aix-Marseille Univ., I2M UMR CNRS 7373\\ 
39 Rue F. Joliot-Curie, 13453 Marseille Cedex 13 - France\\ }
\affil[3]{Univ. Paris-Est, LIGM UMR CNRS 8049\\ 
5 Bd Descartes, 77454 Marne-la-Vall\'ee - France}

\renewcommand\Authands{ and }

\begin{document}

\date{}
\maketitle

\begin{abstract}
Unveiling meaningful geophysical information from seismic data requires to deal with both random and structured ``noises''. As their amplitude may be 
greater than signals of interest (primaries), additional prior information is especially important in performing efficient signal separation. We address here the problem of multiple reflections, caused by  wave-field bouncing between layers. Since 
only approximate models of these phenomena are available, we propose a flexible framework for time-varying adaptive filtering of seismic signals, using sparse representations,  based on inaccurate templates. We recast the joint estimation of adaptive filters and primaries in a new convex variational formulation. This approach allows us to incorporate plausible knowledge about noise statistics, data sparsity and slow filter variation in parsimony-promoting wavelet frames.  The designed primal-dual algorithm solves a  constrained  minimization problem that alleviates standard regularization issues in finding hyperparameters. The approach demonstrates  significantly good performance in low signal-to-noise ratio conditions, both for simulated and real field seismic data.
\end{abstract}

\section{Introduction}
\label{sec:intro}

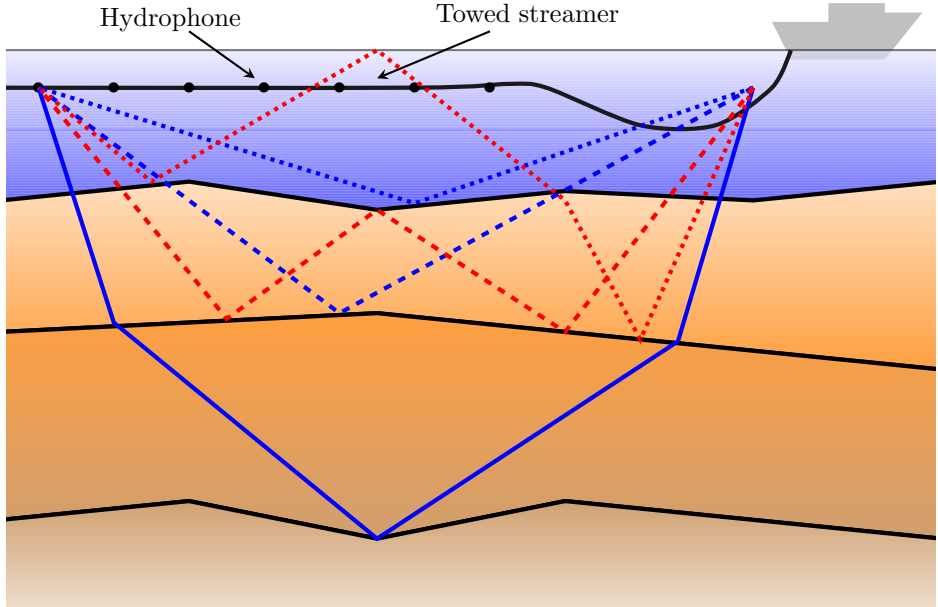
\begin{figure}[htb]
\begin{center}
\begin{tikzpicture}[scale=2.5]
\begin{scope} 
\clip ($(0,0)+2*(\pgflinewidth,\pgflinewidth)$) rectangle ($(5,3.5)-2*(\pgflinewidth,\pgflinewidth)$);
\fill   [xshift=0.2cm, fill=Silver]  (4,2.95) --  (4.5,2.95)  --  (4.7,3.20)  --  (4.50,3.20)  --  (4.50,3.25)  --  (4.20,3.25)  -- (4.20,3.15)  --  (3.9,3.15) -- cycle;
\draw[thick,bottom color=blue!90,top color = blue!10,opacity=.5] (0,2.2) -- (1,2.3) -- (2,2.15) -- (3,2.25) -- (4,2.2) -- (5,2.3) -- (5,3) -- (0,3) -- cycle;
\draw [ultra thick,black!90, xshift=0.2cm] plot [smooth, tension=0.5] coordinates { (4.0,3.0) (3.9,2.8) (3.6,2.6) (3.2,2.6) (2.7,2.8)  (2.5,2.82) (2,2.8) (-0.4,2.8)};
\draw (0.2,2.8) node {$\bullet$} (0.6,2.8) node {$\bullet$} (1.0,2.8) node {$\bullet$} (1.4,2.8) node {$\bullet$} (1.8,2.8) node {$\bullet$} (2.2,2.8) node {$\bullet$} (2.6,2.8) node {$\bullet$};
\draw (2.8,3.1) node[above] {Towed streamer} ; 
\draw [>=stealth,->,thick] (2.6,3.1) -- (2,2.85) ; 
\draw (0.9,3.05) node[above] {Hydrophone} ; 
\draw [>=stealth,->,thick] (1.0,3.1) -- (1.35,2.85) ; 
\draw[ultra thick,bottom color=orange!75,top color = orange!25]  (0,2.2) -- (1,2.3) -- (2,2.15) -- (3,2.25) -- (4,2.2) -- (5,2.3) --  (5,1.3) -- (2,1.6) -- (0, 1.5) --  cycle;
\draw[ultra thick,bottom color=brown!75,top color = orange!75] (0, 1.5) --  (2,1.6) --  (5,1.3) --  (5,0.4) -- (4,0.5) -- (3,0.6) -- (2,0.4) --   (1,0.6) -- (0,0.5) -- cycle;
\draw[ultra thick,bottom color=brown!25,top color = brown!75] (0,0.5) -- (1,0.6) -- (2,0.4) -- (3,0.6) -- (4,0.5) -- (5,0.4) -- (5,0) -- (0,0) -- cycle;
\draw [ultra thick,  color=blue, dotted] (4,2.8) -- (2.2,2.185)-- (0.2,2.8);
\draw [ultra thick,  color=blue, dashed] (4,2.8) -- (1.8,1.6)-- (0.2,2.8);
\draw [ultra thick,  color=blue] (4,2.8) -- (3.6,1.45)-- (2.0,0.4)--  (0.6,1.55)-- (0.2,2.8);
\draw [ultra thick,  color= red!100, dashed] (4,2.8) -- (3.0,1.5)-- (2.0,2.15) -- (1.2,1.57) -- (0.2,2.8);
\draw [ultra thick,  color= red!100, dotted] (4,2.8) -- (3.4,1.46)-- (3,2.2) -- (2,3) -- (0.8,2.3) -- (0.2,2.8);

\end{scope}
\end{tikzpicture}
\end{center}
\vspace{-0.3cm}
\caption{Principles of marine seismic data acquisition and wave propagation. Towed streamer with hydrophones. Reflections on
different layers (primaries in  blue), and  reverberated disturbances (multiple in dotted and dashed red).\label{fig_marine_seismic_reflection_multiple}}
\vspace{-0.4cm}
\end{figure}

Adaptive filtering techniques play a prominent part in signal processing. They cope with time-varying or non-stationary signals and systems. The rationale of these methods is to optimize parameters of variable filters, according to adapted cost functions working on error signals. The appropriate choice of cost functions, that encode a priori information on the system under study, should be balanced with the tractability of the adaptation. While traditional adaptive algorithms resort to least squares minimization, they may be sensitive to outliers, and may not directly promote  simple filters (well-behaved, with concentrated coefficients), especially when the filter length is not well known. 

Certain  systems, for instance transmission channels, behave parsimoniously. They are modeled by sparse impulse response filters with a few large  taps, most of the others being small. Several designs have thus turned toward cost functions promoting filter sparsity  \cite{Duttweiller_D_2000_j-ieee-tsap_proportionate_nlmsaec,Martin_R_2002_j-ieee-tsp_exploiting_saf,Khong_A_2006_p-asilomar_efficient_usaf}. Recently,  developments around   proximity operators \cite{Moreau_J_1962_j-c-r-acad-sci-ser-a-math_fonctions_cdppeh} with signal processing applications \cite{Combettes_P_2011_incoll_proximal_smsp} have allowed  performance improvements. For instance, \cite{Murakami_Y_2010_p-icassp_sparse_afutvstt,Yukawa_M_2012_p-iscas_sparsity-aware_afblpnistt} allow sparsity promotion with $\ell_1$ and $\ell_p$, $0 < p < 1$, quasi-norms, respectively, via time-varying soft-thresholding operators. Improvements reside in convergence speed acceleration  or gains in signal-to-noise ratios (SNRs). These developments are generally performed directly in the signal domain.

Sparsity may additionally be present in signals.  Choosing  an appropriate transformed domain could, when applied appropriately \cite{Gilloire_V_1992_j-ieee-tsp_adaptive_fsbcsaeaaec},  ease the efficiency of adaptive filters \cite{Usevitch_B_1996_j-ieee-tcas2_adaptive_fufb,Hosur_S_1997_j-ieee-tsp_wavelet_tdafirf,Petraglia_M_2008_j-ieee-tsp_nonuniform_safcs}. Such transforms include filter banks \cite{Gauthier_J_2009_j-ieee-tsp_optimization_socfb}  or redundant wavelets \cite{Chaux_C_2007_j-ieee-tit_noise_cpdtwd}. The usefulness of sparsity-promoting loss functions or shrinkage functions in structured data denoising or deconvolution is well documented \cite{Chaux_C_2008_j-ieee-tsp_nonlinear_sbemid,Pesquet_J_2009_j-ieee-tsp_sure_adsidp,Bruckstein_A_2009_j-siam-review_sparse_ssesmsi}. 
Geophysical signal processing \cite{Robinson_E_1980_book_geophysical_sa} is a field where dealing  with sparsity, or at least energy concentration, both in the system filter and the data domain, is especially beneficial.

The aim of seismic data analysis is to infer  the subsurface structure from seismic  wave fields  recorded through land or marine acquisitions. In reflection seismology,  seismic waves, generated by a close-to-impulsive source, propagate through the subsurface medium. They travel  downwards, then upwards, reflected by geological interfaces,  convolved  by earth filters. They account for the unknown relative distances and velocity contrasts between layers and they are affected by  propagation-related distortions. A portion of the wave fields  is finally recorded near the surface by arrays of seismometers (geophones or hydrophones). In marine acquisition, hydrophones are towed by kilometer-long streamers.

Signals of interest, named primaries, follow wave paths  depicted in  dotted, dashed and solid blue  in Fig.  \ref{fig_marine_seismic_reflection_multiple}.
Although the contributions are generally considered linear, several types of disturbances, structured or more stochastic, affect the relevant information  present in seismic data. Since the data recovery problem is under-determined, geophysicists have developed pioneering sparsity-promoting techniques. For instance, robust, $\ell_1$-promoted deconvolution 
\cite{Claerbout_J_1973_j-geophysics_robust_med}  
 or  complex  wavelet transforms 
\cite{Morlet_J_1975_p-seg_seismic_tiqm} 
still pervade many areas of signal processing. 

We address one of the  most severe types of interferences: secondary reflections, named multiples, corresponding to seismic waves bouncing between  layers \cite{Essenreiter_R_1998_j-ieee-tsp_multiple_rasdb}, as illustrated with red dotted  and dashed lines  in Fig.  \ref{fig_marine_seismic_reflection_multiple}. These reverberations  share  waveform and frequency contents similar to  primaries, with longer propagation times. 
 From the standpoint of geological information interpretation, they often  hide deeper target reflectors. For instance, the dashed-red multiple path may possess a total travel time comparable with that of the solid-blue primary. Their separation is thus required for accurate subsurface characterization. A geophysics industry standard  consists of model-based multiple filtering. One or several realistic templates of a potential  multiple are determined off-line, based on primary reflections identified in above layers. 
For instance, the dashed-red path may be approximately  inferred from the dashed-blue, and then adaptively filtered for separation from the solid-blue propagation.  Their precise estimation is beyond the scope of this work, we suppose them given by prior seismic processing or modeling.  As  template modeling is partly  inaccurate --- in delay, amplitude and frequency --- templates should be adapted in a time-varying fashion before being subtracted from the recorded data. Resorting  to several templates and  weighting them adaptively,  depending on the time and space location of   seismic traces, helps when highly complicated  propagation paths occur. Increasing the number of templates is a growing trend in exploration. Meanwhile, inaccuracies in template modeling, complexity  of  time-varying adaptation  combined with additional  stochastic disturbances require additional constraints to obtain geophysically-sound solutions. 

We propose a methodology for primary/multiple  adaptive separation based on approximate templates. This framework addresses at the same time  structured reverberations and a more stochastic part. Namely, let  $n\in \{0,\ldots,N-1\}$ denote the time index for the observed seismic trace $z$, acquired by a given sensor. We assume, as customary in seismic, an additive model of contributions:
\begin{equation}
z^{(n)} = \overline{y}^{(n)} + \overline{s}^{(n)} +  b^{(n)}\,.
\label{eq:model}
\end{equation} 
The unknown signal of interest (primary, in blue) and the sum of undesired, secondary reflected signals (different multiples, in red) are denoted, respectively, by $\overline{y}=(\overline{y}^{(n)})_{0 \le n < N}$ and $\overline{s}=(\overline{s}^{(n)})_{0 \le n < N}$. Other unstructured contributions are gathered in the noise term  $b= (b^{(n)})_{0 \le n < N}$. We assume that several approximate templates accounting for multiples are available. 
As the above problem is undetermined, additional constraints should be devised.  We specify sparsity and slow-variation requirements on  primaries and adaptive filters. In Section~\ref{sec:related-work}, we analyze related works and specify the novelty of the proposed methodology. To the authors' knowledge, the formulation of this template-based restoration problem in a nonstationary context, taking into account noise, sparsity, slow adaptive filter variation, along with constraints on filters is unprecedented, especially in the field of seismic processing.  Section \ref{sec:model} describes  the transformed linear model  incorporating the templates with adaptive filtering. In Section \ref{sec:var-formulation}, we  formulate a  generic variational form for the problem. Section \ref{sec:algo}
describes the primal-dual proximal formulation. The performance of the proposed method is assessed in Section \ref{sec:results}. We detail the chosen optimization criteria and provide a  comparison with different types of frames. The methodology is first evaluated on a realistic synthetic data model, and finally tested and applied to an actual seismic data-set.
Conclusions and perspectives are drawn in Section~\ref{sec:conclusion}. This work improves upon  \cite{Gragnaniello_D_2012_p-eusipco_convex_vamrsd} by  taking into account several multiple templates. Part of it was briefly presented in \cite{Pham_M_2013_p-icassp_seismic_mrpdpa}, by incorporating  an additional  noise  into the generic model, and by introducing  alternative norms  in  multiple selection objective criteria. Here, the approach is extended. In particular, the problem is completely reformulated as a constrained minimization problem, in order to simplify the determination of data-based parameters, as compared with our previous regularized approach involving hyper-parameters.

\section{Related and proposed work}
\label{sec:related-work}
Primary/multiple separation is a long standing problem in seismic. Published solutions are weakly generic, and often embedded in a more general processing work-flow. Levels of prior knowledge --- from the shape of the seismic source to partial geological information ---    greatly  differ depending on  data-sets. We  refer to  \cite{Weglein_A_2011_j-tle_multiple_arara2011,Ventosa_S_2012_j-geophysics_adaptive_mswbcuwf} for recent accounts on broad processing issues, including shortcomings of standard $\ell_2$-based methods. The latter are computationally efficient, yet their performance decreases when traditional assumptions fail (primary/multiple decorrelation, weak linearity or stationarity, high noise levels).
We focus here on recent sparsity-related approaches, pertaining to geophysical signal processing. The potentially parsimonious layering of the subsurface (illustrated in Fig.  \ref{fig_marine_seismic_reflection_multiple}) suggests a modeling of primary reflection coefficients with generalized Gaussian or Cauchy distributions 
\cite{Walden_A_1986_j-geophys-prospect_nature_ngprcsd}, 
having suitable  parameters. The  sparsity induced on seismic data has influenced deconvolution and multiple subtraction. Progressively, the non-Gaussianity of seismic traces has been emphasized, and contributed to the use of  more robust norms 
\cite{Duarte_L_2012_p-eusipco_seismic_wsmrpca,Takahata_A_2012_j-ieee-spm_unsupervised_pgsrskabdbss} 
for blind separation with independent component analysis (ICA) for the signal of interest. As the true nature of seismic data distribution is still debated, 
including its stationarity \cite{Bois_P_1963_j-geophys-prospect_etude_scmssr}, 
a handful of works have investigated processing in appropriate transformed domains. They may either stationarize \cite{Krim_H_1995_j-ieee-tit_multiresolution_acnp} or strengthen data sparsity. For instance, \cite{Donno_D_2011_j-geophysics_improving_mrulsdfica} applies ICA in a  dip-separated domain. In \cite{Neelamani_R_2010_j-geophysics_adaptive_sucvct}, as well as in \cite{Herrmann_F_2004_p-cseg_curvelet_ipame} and subsequent works by the same group, a special focus is laid on separation in the curvelet domain. 

Aforementioned works mostly  deal with the mitigation of some $\ell_2$-norm on residuals, as remnant noise is traditionally considered Gaussian in seismic. They are blended   with  $\ell_0$ objectives, solved through  $\ell_1$ or hybrid  $\ell_1$-$\ell_2$ approximations \cite{Guitton_A_2004_j-geophys-prospect_adaptive_smul1n}, resorting for instance to iteratively re-weighted least-squares method. Recently, \cite{Costagliola_S_2011_p-eage_hybrid_nasmr} investigated the use of intermediate $\ell_p$-norms, with $p=1.2$ for instance, accounting for the ``super-Gaussian nature of the seismic data due to the interfering fields'', in the time domain. Without further insights on  precise modeling, a more flexible framework  is desirable to adapt to the nature of different seismic data, either in the direct or in a variety of transformed domains.

Data sparsity and noise Gaussianity alone may not be sufficient to solve \eqref{eq:model}. Additional constraints reduce the set of solutions, hopefully to   geologically sounder ones. A first one is the locality of  matched filters, traditional in standard multiple filtering. These can be modeled by Finite Impulse Response (FIR) operators. Classical filter support limitations, down to one-tap \cite{Neelamani_R_2010_j-geophysics_adaptive_sucvct,Ventosa_S_2012_j-geophysics_adaptive_mswbcuwf}, assorted with $\ell_2$ or $\ell_1$ criteria, are standard. In other seismic processing fields, \cite{Gholami_A_2012_j-ieee-tgrs_fast_asdpo} has investigated mixed $\ell_p$-$\ell_1$ loss functions for deconvolution.
Recently, in
\cite{Yang_Y_2012_tr_seismic_drmc,Kreimer_N_2013_p-icassp_nuclear_nmtces} 
the use of the nuclear norm is promoted for interpolation, combined with a standard $\ell_2$-norm penalty.  Yet, to the authors' knowledge, no   work  in multiple removal has endeavored a more systematic study of  variational and sparsity constraints on the adaptive filters, in the line of \cite{OBrien_M_1994_j-ieee-tsp_recovery_sstsl1nd}.  In this work, we propose a formulation allowing a family of penalties to be applied to the adaptive FIR filters. Since no metric is evidently more natural, such a flexibility is useful to assess different objectives. For instance, one might be interested in either well-preserving primaries, in mild noise cases, or robustly removing the multiples, in high contamination situations. Indeed, when the perturbation is stronger in amplitude than the target signal, geophysicists are interested in uncovering even spoors of potential primaries, obfuscated by noise. As will be seen, the most appropriate norm depends on such contexts.
Finally, the propagation medium, as well as the modeled templates,  carry continuous variations. With the seismic bandwidth (up to \SI{125}{\hertz}), changes in  signals are not as dramatic as in sharp images. Consequently, we expect the adapted filters to exhibit bounded variations from one time index to the next one.

This paper presents for the first time a relatively generic framework for multiple reflection filtering with (\textit{i}) a noise prior, (\textit{ii}) sparsity constraints on signal frame coefficients, (\textit{iii}) slow variation modelling of the adaptive filters, and (\textit{iv}) concentration metrics on the filters. With the development of recent optimization tools, multiple constraints can now be handled in a convenient manner. 
 Due to the diversity of focus points, paired with data observation, we choose here to decouple effects and  to insist on  (\textit{iv}), with respect to different flavors of 1D wavelet bases and frames \cite{Pesquet_J_1996_j-ieee-tsp_time-invariant_owr,Fowler_J_2005_j-ieee-spl_redundant_dwtan}, which  appear as  natural atoms for sparse
descriptions of some physical processes, related to propagation
and reflection of signals through media. 

The evaluation of the proposed multiple filtering algorithm on seismic data is not straightforward, for two main reasons. First,   seismic processing work-flows  are neither publicly available for benchmarks and are generally heavily parametrized. Second, quality measures are not easy to devise since visual inspection is of paramount importance in geophysical data processing assessment. We thus compare the proposed approach with a state-of-the-art solution, previously benchmarked against industrial competitors \cite{Ventosa_S_2012_j-geophysics_adaptive_mswbcuwf}.

\section{Model description}
\label{sec:model}
We assume that multiple templates are modeled at the temporal vicinity of actual disturbances, with  standard geophysical assumptions on primaries. The multiple signal possesses a local behavior related to the geological context. Hence, we assume the availability of 
  $J$ templates $(r^{(n)}_j)_{0 \le n < N, 0 \le j < J}$,  related to $(\overline{s}^{(n)})_{0\le n< N}$ via a possibly non-causal linear model through a limited support relationship:
\begin{equation}\label{e:sn}
\overline{s}^{(n)} = \sum_{j=0}^{J-1}\, \sum_{p=p'}^{p'+P_j-1} \overline{h}_j^{(n)}(p) r_j^{(n-p)}
\end{equation}
where $\overline{h}_j^{(n)}$ is an unknown finite impulse response (with $P_j$ tap coefficients) associated with template $j$ and time $n$, and where $p' \in \{-P_j+1,\ldots,0\}$ is its starting index ($p'=0$ corresponds to the causal case). It must be emphasized that the dependence w.r.t. the time index $n$ of the impulse responses implies that the filtering process is time variant, although it can be assumed slowly varying in practice. Indeed, seismic waveforms evolve gradually with propagation depth, in contrast with  steeper variations around contours in natural images.
Templates are generated with standard geophysical modeling based on the above primaries. The adaptive FIR assumption is commonly adopted, and applied in partly overlapping, complementary time windows at different scales. The observation that adapted filters are ill-behaved, due to the band-pass nature of seismic data is well known, although rarely documented, motivating the need for filter coefficient control.
Defining vectors $\overline{s}$ and $(\overline{h}_j)_{0 \le j < J}$  by:
\begin{align}
\overline{s} &= \begin{bmatrix} \overline{s}^{(0)}& \cdots & \overline{s}^{(N-1)}\end{bmatrix}^\top,\nonumber\\
\overline{h}_j &= \left[ \overline{h}_j^{(0)}(p') \; \cdots \; \overline{h}_j^{(0)}(p'+P_j-1) \; \cdots 
  \overline{h}_j^{(N-1)}(p') \; \cdots \;  \overline{h}_j^{(N-1)}(p'+P_j-1)\right]^\top,\nonumber
\end{align}%
and  block diagonal matrices $(R_j)_{0 \le j < J}$   of size $N \times NP_j$:
\begin{equation}\nonumber
R_j = \begin{bmatrix} 	R_j^{(0)} 	& 0	     & \ldots & 0	\\
0 	& R_j^{(1)}	     & \ldots & 0	\\
\vdots		& 0	     & \ddots & \vdots		\\
0	& 0	     & \ldots & R_j^{(N-1)}
\end{bmatrix},
\end{equation}
where $(R_j^{(n)})_{0\le n \le N-1}$ are vectors of dimension $P_j$ such that
\begin{align}
&\left[(R_j^{(0)})^\top (R_j^{(1)})^\top \cdots (R_j^{(N-1)})^\top\right]^\top=
&{\small
\begin{bmatrix}
r_j^{(-p')} & \cdots & r_j^{(0)} & 0 & \cdots & &  0 \\
r_j^{(-p'+1)} & \cdots &  &  r_j^{(0)} & 0 & \cdots & 0 \\
\vdots  &  &  &  & \\
r_j^{(N-1)} & r_j^{(N-2)}  &  & \cdots &  &  & r_j^{(N-P_j)} \\
0 & r_j^{(N-1)} &  & \cdots  &  &  & r_j^{(N-P_j+1)}\\
\vdots  &  &  &  &  & \\
0 & \cdots & 0 & r_j^{(N-1)} & \cdots & &  r_j^{(N-P_j-p')}\\
\end{bmatrix}.}
\nonumber
\end{align}
Eq. \eqref{e:sn} can be expressed more concisely as
\begin{equation}
\overline{s} = \sum_{j=0}^{J-1} R_j \overline{h}_j\,.\nonumber
\end{equation}
For more conciseness, one can write $\overline{s}=\bR\overline{\bh}$ 
by defining $\bR=[R_0 \ldots R_{J-1}] \in \RR^{N \times Q}$ where $Q=NP$ with $P=\sum_{j=0}^{J-1} P_j$ and $\overline{\bh}=[\overline{h}_0^\top \ldots \overline{h}_{J-1}^\top]^\top \in \RR^Q$.

\section{Variational formulation of the problem}
\label{sec:var-formulation}

\subsection{Bayesian framework}
\label{ssec:Bayes}

We assume that the characteristics of the primary are appropriately described through a prior statistical model in 
a (possibly redundant) frame of signals, e.g. a wavelet frame \cite{Christensen_O_2008_book_frames_bic}.
If we denote by $\overline{x}$ the vector of frame coefficients, and $F\in \RR^{K\times N}$ designates the associated analysis operator, we have \cite{Chaux_C_2007_j-inv-prob_variational_ffbip}: $\overline{x} = F \overline{y}$.
In addition, we assume that $\overline{y}$ is a realization of a random vector $Y$, the probability density function (pdf) of which is given by
\begin{equation}\label{e:priorfY}
(\forall y \in \RR^N)\quad
f_Y(y) \propto \exp(-\varphi(Fy))
\end{equation}
where $\varphi\colon \RR^N \to \RX$ is the associated potential, assumed to have a fast enough decay.

On the other hand, to take into account the available information on the unknown filter, 
it can be assumed that for all $j\in\{0,\ldots,J-1\}$, $\overline{h}_j$ is a realization of a random vector $H_j$. 
Let $\hFilterSet =  \RR^{NP_0}\times \cdots \times \RR^{NP_{J-1}}$.
The joint pdf of the filter coefficients can be expressed as:
\begin{equation}
(\forall h \in \hFilterSet)\quad f_{H_0,\ldots,H_{J-1}}(h) \propto \exp(-\rho(h)),\nonumber
\end{equation}
where $(H_0,\ldots,H_{J-1})$ is independent of $Y$.
It is further assumed that the noise vector $b$ is a realization of a random vector $B$ with pdf
\begin{equation}
(\forall b \in \RR^{N})\qquad
f_B(b) \propto \exp(-\psi(b)),\nonumber
\end{equation}
where $\psi : \RR^N \to \RX$, and that $B$ is independent of $Y$ and $H_0,\ldots,H_{J-1}$.
The posterior distribution of $(Y,H_0,\ldots,H_{J-1})$ conditionally to $Z = Y + \sum_{j=0}^{J-1}R_jH_j+B$ is then given by
\begin{multline}
(\forall y\in \RR^N) (\forall h \in \hFilterSet)
f_{Y,H_0,\ldots,H_{J-1}\mid Z = z}(y,h) 
\propto \exp\left(-\psi\Big(z-y-\sum_{j=0}^{J-1}R_jh_j\Big)\right)\nonumber
 f_{Y}(y) f_{H_0,\ldots,H_{J-1}}(h).
\end{multline}

By resorting to an estimation of $(\overline{y},\overline{h}_0,\ldots,\overline{h}_{J-1})$ in the sense of the MAP (Maximum A Posteriori), the problem can thus be formulated under the following variational form:
\begin{equation}
\underset{y\in \RR^N,\,h \in \hFilterSet}{\text{minimize}}\; \psi(z-y-\sum_{j=0}^{J-1}R_jh_j)+\varphi(F y) + \rho(h).\nonumber
\end{equation}

\subsection{Problem formulation}
For simplicity, we propose to adopt uniform priors for $Y$ and $(H_0,\ldots,H_{J-1})$ by choosing for $\varphi$
and $\rho$ indicator functions of closed convex sets.
The associated MAP estimation problem then reduces to 
the following constrained minimization problem:
\begin{align}
\underset{Fy\in D,\, \bh \in C}{\text{minimize}}\; \Psi\left(\begin{bmatrix}
y\\ \bh
\end{bmatrix}\right) 
\label{eq:fct_constrain}
\end{align}
\noindent where the data fidelity term is defined by function
$\Psi\colon \left(\begin{bmatrix}
y\\ \bh
\end{bmatrix}\right) \mapsto \psi\left(z-[\I\;\bR]\begin{bmatrix}
y\\ \bh
\end{bmatrix}\right)$,
and the a priori information available on the filters and the primary are expressed through hard constraints modeled by nonempty closed convex sets $C$ and $D$.
One of the potential advantages of such a constrained formulation is that it facilitates the choice of the related parameters with respect to the regularized approach which was investigated in some of our previous works \cite{Gragnaniello_D_2012_p-eusipco_convex_vamrsd,Pham_M_2013_p-icassp_seismic_mrpdpa} (this point will be detailed later on).
We will  now turn our attention to the choice of $\Psi$, $C$ and $D$.

\subsection{Considered data fidelity term and constraints}
\label{ssec:dataFid_const}

\subsubsection{Data fidelity term}
\label{ssec:fidel}
Function $\Psi$ accounts for the noise statistics. In this work, the noise is assumed to be additive, zero-mean, white and Gaussian. This leads to the quadratic form $\psi=\| \, . \,\|^2$.

\subsubsection{A priori information on $\bh$}
The filters are assumed to be time varying. However, in order to ensure smooth variations along time, we propose to introduce
constraint sets
\begin{multline}
C_1 = \Big\{ \bh \in \RR^{Q} \mid \forall (j,p), \forall n \; \in \left\{0,\dots,\left\lfloor\frac{N}{2}\right\rfloor-1\right\}
  |h_j^{(2n+1)}(p) - h_j^{(2n)}(p)| \le \varepsilon_{j,p} \Big\}
  \label{eq:C_1}
\end{multline}
\begin{multline}
C_2 = \Big\{ \bh \in \RR^{Q} \mid \forall (j,p), \forall n \; \in 
\left\{1,\dots,\left\lfloor\frac{N-1}{2}\right\rfloor\right\}
  |h_j^{(2n)}(p) - h_j^{(2n-1)}(p)| \le \varepsilon_{j,p} \Big\}.
  \label{eq:C_2}
\end{multline}
These constraints prevent strong variations of corresponding coefficients of the impulse response, estimated at two consecutive times. The bounds $\varepsilon_{j,p} \in [0,\pinf[$ may depend on the shape of the expected filter. For example, its dependence on the coefficient index $p$ may enable a larger (resp. smaller) difference for filter coefficients taking larger (resp. smaller) values.
Moreover, some additional a priori information can be added directly on the vector of filter coefficients $\bh$. This amounts to defining a new convex set $C_3$ as a lower level set of some lower-semicontinuous convex function $\widetilde{\rho}$, by setting
$C_3=\left\{\bh \in \RR^{Q} \mid \widetilde{\rho}(\bh) \leq \lambda\right\}$ where $\lambda \in \RPP$.
$\widetilde{\rho}\colon \RR^{Q} \to [0,+\infty[$ may correspond to simple norms such as $\ell_1$ or $\ell_2$-norms but also to more sophisticated ones such as a mixed $\ell_{1,2}$-norm \cite{Kowalski_M_2009_j-acha_sparse_rumn}.
Hence, the convex set $C$ is defined as $C=C_1 \cap C_2  \cap C_3$. From a computational standpoint (see Section \ref{sec:algo}), it is more efficient to split $C$ into three subsets as described above.

\subsubsection{A priori information on $y$}
As mentioned in Section~\ref{ssec:Bayes}, we assume that the primary signal is sparsely described through an analysis frame operator $F\in \RR^{K\times N}$ \cite{Jacques_L_2011_j-sp_panorama_mgrisdfs}, which may ease its processing, by increasing the data-domain discrepancy between primaries and multiples. 
The associated constraint can be split by defining a partition of $\{1, \ldots ,K\}$ denoted by $\{\KK_\ell \mid \ell \in \{1,\ldots,\mathcal{L}\}\}$. For example, for wavelet frames, $\mathcal{L}$ may correspond to the number of subbands and $\KK_\ell$
is the $\ell$-th subband.
Then, one can choose $D = D_1 \times \cdots \times D_{\mathcal{L}}$
with $D_{\ell} = \{ (x_k)_{k\in \KK_\ell} \mid \sum_{k\in \KK_\ell} \widetilde{\varphi}_\ell(x_k) \leq \beta_\ell \}$, where, for every $\ell \in \{1,\ldots,\mathcal{L}\}$, $\beta_\ell \in \RPP$, and $\widetilde{\varphi}_\ell : \RR \to [0,+\infty[$ is a lower-semicontinuous convex function.

\section{Primal-Dual proximal algorithm}
\label{sec:algo}

Our objective  is to provide a numerical solution to Problem \eqref{eq:fct_constrain}. This amounts to minimizing function $\Psi$ with respect to $y$ and $\bh$, the latter variables being constrained to belong to the constraint sets $D$ and $C$, respectively. These constraints are expressed through linear operators, such as a wavelet frame analysis operator $F$. For this reason, 
primal-dual algorithms \cite{BricenoArias_L_2011_j-siam-optim_monotone_pssmcmid,Chambolle_A_2011_j-math-imaging-vis_first_opdacpai,Condat_L_2013_j-optim-theory-appl_primal-dual_smcoilplct}, such as the Monotone+Lipschitz Forward-Backward-Forward (M+L FBF) algorithm \cite{Combettes_P_2012_j-set-valued-var-anal_primal_dsasimclpstmo}, constitute appropriate choices since they avoid some large-size matrix inversions inherent to other schemes such as the ones proposed in  \cite{Afonso_M_2011_j-ieee-tip_augment_lacofiip,Pesquet_J_2012_j-pacific-j-opt_parallel_ipom}.
As mentioned in Section \ref{ssec:fidel}, $\Psi$ is a quadratic function and its gradient is thus Lipschitzian, which
allows it to be directly handled in the M+L FBF algorithm.
In order to deal with the constraints, projections onto the closed convex sets $(C_m)_{1\le m \le 3}$ and $D$ are  performed (these projections are  described in more details in the next section).

\subsection{Gradient and projection computation}
\label{sec:choice}

From the assumption of additive zero-mean Gaussian noise, we deduce that $\Psi$ is differentiable with a $\mu$-Lipschitzian gradient, i.e.
$(\forall \begin{bmatrix}
y\\ \bh
\end{bmatrix} \in \RR^{N+Q}) (\forall \begin{bmatrix}
y^{'}\\ \bh^{'}
\end{bmatrix}\in \RR^{N+Q})$:
$$ \| \nabla\Psi\Big(\begin{bmatrix}
y\\ \bh
\end{bmatrix}\Big) -\nabla\Psi
\Big(\begin{bmatrix}
y^{'}\\ \bh^{'}
\end{bmatrix}\Big) \| \leq \mu \| \begin{bmatrix}
y\\ \bh
\end{bmatrix} - \begin{bmatrix}
y^{'}\\ \bh^{'}
\end{bmatrix} \|$$
and
\begin{equation}
\nabla \Psi =2[\I\;\bR]^\top([\I\;\bR] \cdot-z).\nonumber
\end{equation}
The gradient of $\Psi$ is thus $\mu$-Lipschitzian with 
\begin{equation}\label{e:constLip}
\mu=2\ns{[\I\;\bR]}^2
\end{equation} 
where $\ns{\cdot}$ denotes the spectral norm.
Note that the proposed method could be applied to other functions $\psi$ than a quadratic one, provided that they are Lipschitz differentiable.

\begin{figure}[htb]
\centering
\includegraphics[width=9cm]{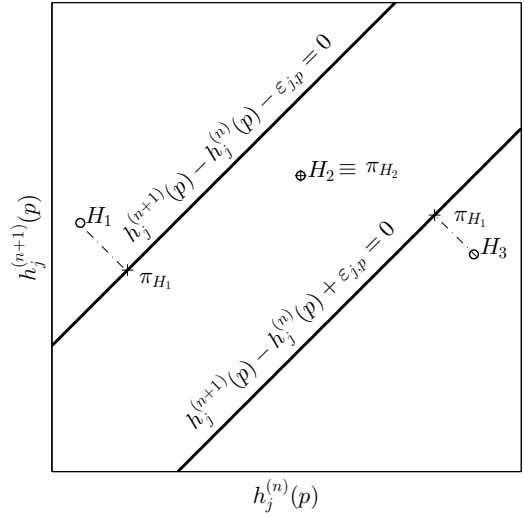}
\caption{\label{fig_projC12}{Projection onto $C_1$/$C_2$ of points $H_1$, $H_2$ and $H_3$ in $\RR^2$.} }
\end{figure}

Now, we turn our attention to the constraint sets $C$ and $D$.
$C$ models the constraints we set on the filters $\bh$, which are split into 3 terms (see Section \ref{ssec:dataFid_const}). We  thus have to project onto each set $C_m$ with $m \in \{1,2,3\}$. The projections onto the two first constraint sets
$C_1$ and $C_2$ --- imposing smooth variations along time of the corresponding tap coefficients --- reduce to projections onto a set of hyperslabs of $\RR^2$ as illustrated in Fig.~\ref{fig_projC12}.

More precisely, the projection onto $C_1$ (the projection onto $C_2$ yielding similar expressions) is calculated as follows:
let $\bh \in \RR^{Q}$ and let $\mathbf{g}_1=\Pi_{C_1}(\bh)$; 
then for every $j \in \{0,\cdots,J-1\}$, $p \in \{p',\cdots,p'+P-1\}$ and
$n \in \left\{0,\dots,\left\lfloor\frac{N}{2}\right\rfloor-1\right\}$,
\begin{enumerate}
 \item if $|h_j^{(2n+1)}(p) - h_j^{(2n)}(p)|<\varepsilon_{j,p}$, then
\begin{align*}
  g_{j,1}^{(2n)}(p)=h_j^{(2n)}(p),\quad g_{j,1}^{(2n+1)}(p)=h_j^{(2n+1)}(p);
\end{align*} 
 \item if $h_j^{(2n+1)}(p) - h_j^{(2n)}(p)>\varepsilon_{j,p}$, then
\begin{align*}
  g_{j,1}^{(2n)}(p)&=\dfrac{h_j^{(2n+1)}(p)+h_j^{(2n)}(p)}{2}-\dfrac{\varepsilon_{j,p}}{2}\\
  g_{j,1}^{(2n+1)}(p)&=\dfrac{h_j^{(2n+1)}(p)+h_j^{(2n)}(p)}{2}+\dfrac{\varepsilon_{j,p}}{2};
\end{align*}
\item if $h_j^{(2n+1)}(p) - h_j^{(2n)}(p)<-\varepsilon_{j,p}$, then
\begin{align*}
  g_{j,1}^{(2n)}(p)&=\dfrac{h_j^{(2n+1)}(p)+h_j^{(2n)}(p)}{2}+\dfrac{\varepsilon_{j,p}}{2}\\
  g_{j,1}^{(2n+1)}(p)&=\dfrac{h_j^{(2n+1)}(p)+h_j^{(2n)}(p)}{2}-\dfrac{\varepsilon_{j,p}}{2}.
\end{align*}
\end{enumerate}

$C_3$ introduces a priori information on the filter vector $\bh$ through the lower-semicontinuous convex function $\widetilde{\rho}$. This function can be chosen separable w.r.t. $j \in \{0,\ldots,J-1\}$ in the sense that
\begin{equation}
\widetilde{\rho}(\bh)=\sum_{j=0}^{J-1} \widetilde{\rho}_j(h_j).\nonumber
\end{equation}
This term can be seen as a concentration measure for the filter tap amplitude. Subsequently, we consider three possible choices for $\widetilde{\rho}_j$:
\begin{enumerate}
\item $\ell_1$-norm: 
\begin{equation}
\widetilde{\rho}_j(h_j)=\|h_j\|_{\ell_1}=\sum_{n=0}^{N-1}\sum_{p=p'}^{p'+P_j-1} |h_j^{(n)}(p)|.\nonumber
\end{equation}
 This choice requires to perform projections onto an $\ell_1$-ball. This can be achieved by using the iterative procedure proposed in \cite{VanDenBerg_E_2008_j-siam-sci-comp_probing_pfbps}, which yields the projection in a finite number of iterations.
\item squared $\ell_2$-norm: 
\begin{equation}
\widetilde{\rho}_j(h_j)=\|h_j\|_{\ell_2}^2=\sum_{n=0}^{N-1}\sum_{p=p'}^{p'+P_j-1} |h_j^{(n)}(p)|^2. \nonumber
\end{equation}
In this case, the projection is straightforward.
\item mixed $\ell_{1,2}$-norm: 
\begin{equation}
\widetilde{\rho}_j(h_j)=\|h_j\|_{\ell_{1,2}}=\sum_{n=0}^{N-1}\left(\sum_{p=p'}^{p'+P_j-1} |h_j^{(n)}(p)|^2\right)^{1/2}. \nonumber
\end{equation}
Then, we can use an algorithm similar to  \cite{VanDenBerg_E_2008_j-siam-sci-comp_probing_pfbps} computing the projection onto an $\ell_1$ ball.
\end{enumerate}

Finally, as mentioned earlier in Section \ref{ssec:dataFid_const}, 
the prior information on the primary $y$ is expressed through the frame analysis operator $F$
by splitting the constraint into individual subband constraints.
In order to promote sparsity of the coefficients, the potential function employed for the $\ell$-th subband 
with $\ell \in \{1,\ldots,\mathcal{L}\}$
can be chosen equal to $\widetilde{\varphi}_\ell = |\cdot|$.
For computing the resulting projection $\Pi_{D_\ell}$ onto an $\ell_1$-ball, we can again employ
 the iterative procedure proposed in \cite{VanDenBerg_E_2008_j-siam-sci-comp_probing_pfbps}.

\subsection{M+LFBF algorithm}
The primal-dual approach chosen to solve the minimization problem \eqref{eq:fct_constrain} is detailed in Algorithm \ref{algo:Primal-Dual}.
It alternates the computations of the gradient of $\Psi$, and of the projections onto $(C_m)_{1\le m \le 3}$ and $(D_\ell)_{1\le \ell \le \mathcal{L}}$.

The choice of the step size is crucial for the convergence speed and it has to be chosen carefully.
First, the norm of each linear operator involved in the criterion or at least an upper bound of it must be available.
In our case, we have:
\begin{equation}
  \ns{[\I \; \bR]} \leq \sqrt{1+\ns{R_0}^2 + \cdots + \ns{R_{J-1}}^2} 
\end{equation}
where $  \ns{R_j} = \max_{n\in\{0,\ldots,N-1\}}\|R_j^{(n)}\|$ for every $j\in \{0,\ldots,J-1\}$.
Secondly, the step size $\gamma^{[i]}$ at each iteration $i$ must be chosen so as to satisfy the following rule: let 
$\mu$ be the Lipschitz constant defined in \eqref{e:constLip}, let
$\beta=\mu+\sqrt{\ns{F}^2+3}$ and let $\epsilon \in ]0,\frac{1}{\beta+1}[$, 
then $\gamma^{[i]} \in [\epsilon,\frac{1-\epsilon}{\beta}]$.  
$\ns{F}^2$ can be easily evaluated. Indeed, in the case of a tight frame, it is equal to the frame constant
and, otherwise, it can be computed by an iterative approach \cite[Algorithm 4]{Chaari_L_2009_p-spie-w_solving_ipotcot}.
It is important to emphasize that the convergence of this algorithm to an optimal solution to Problem \eqref{eq:fct_constrain} 
is guaranteed by \cite[Theorem 4.2]{Combettes_P_2012_j-set-valued-var-anal_primal_dsasimclpstmo}.
In practice, the higher the norms of $F$ and $(R_j)_{0\le j \le J-1}$, the slower the convergence of the algorithm. In order to circumvent this difficulty, one can resort to a preconditioned version of the algorithm \cite{Repetti_A_2012_p-eusipco_penalized_wlsardcsdn}. However, this was not found to be useful in our experiments.

\begin{algorithm}[!tbh]
\caption{Primal-dual algo. M+LFBF to solve \eqref{eq:fct_constrain}\label{algo:Primal-Dual}}
\begin{algorithmic}
\STATE Let $\gamma^{[i]} \in [\epsilon,\frac{1-\epsilon}{\beta}]$
\STATE Let $\begin{bmatrix}
y^{[0]}\\
\bh^{[0]}
\end{bmatrix} \in \RR^{N+Q},\; v^{[0]} \in \RR^K, \left(u_m^{[0]}\right)_{m\in\{1,2,3\}} \in (\RR^{Q})^3\, \; s_2^{[0]} \in \RR^K, \; w_1^{[0]} \in \RR^K$
\FOR{$i=0,1,\dots$}
\STATE \textsl{Gradient computation}
\STATE {\small $\begin{bmatrix}
s_{1}^{[i]}\\
t_{1}^{[i]}
\end{bmatrix}
=  \begin{bmatrix}
y^{[i]}\\
\bh^{[i]}
\end{bmatrix}
- \gamma^{[i]}\left(\nabla\Psi
\Big(\begin{bmatrix}
y^{[i]}\\
\bh^{[i]}
\end{bmatrix}\Big) + \begin{bmatrix}
F^* v^{[i]}\\
\sum_{m=1}^3 u_m^{[i]}
\end{bmatrix}\right)
$}
\STATE \textsl{Projection computation}
\STATE $x_1^{[i]} = Fy^{[i]}$
\FOR{$\ell=1:\mathcal{L}$}
\STATE $\left(s_{2}^{[i]}(k)\right)_{k \in \KK_\ell}=\left(v^{[i]}(k)+\gamma^{[i]} x_1^{[i]}(k)\right)_{k \in \KK_\ell}$ 
\STATE {\small $\left(w_{1}^{[i]}(k)\right)_{k \in \KK_\ell} = \left(s_{2}^{[i]}(k)\right)_{k \in \KK_\ell}-\gamma^{[i]} \Pi_{D_\ell}\left(\frac{\left(s_{2}^{[i]}(k)\right)_{k \in \KK_\ell}}{\gamma^{[i]}}\right)$ }
\ENDFOR
\FOR{$m=1:3$}
\STATE $t_{2,m}^{[i]}=u_m^{[i]}+\gamma^{[i]} \bh^{[i]}$
\STATE $w_{2,m}^{[i]} = t_{2,m}^{[i]}-\gamma^{[i]} \Pi_{C_m}\left(\frac{t_{2,m}^{[i]}}{\gamma^{[i]}}\right)$
\ENDFOR
\STATE \textsl{Averaging}
\STATE $x_{2}^{[i]} = F s_{1}^{[i]}$ 
\FOR{$\ell=1:\mathcal{L}$}
\STATE $\left(q_{1}^{[i]}(k)\right)_{k \in \KK_\ell}= \left(w_{1}^{[i]}(k)+\gamma^{[i]} x_{2}^{[i]}(k)\right)_{k \in \KK_\ell}$
\STATE $\left(v^{[i+1]}(k)\right)_{k \in \KK_\ell} = \left(v^{[i]}(k) - s_{2}^{[i]}(k) + q_{1}^{[i]}(k)\right)_{k \in \KK_\ell}$
\ENDFOR
\FOR{$m=1:3$}
\STATE $q_{2,m}^{[i]} = w_{2,m}^{[i]}+\gamma^{[i]} t_{1}^{[i]}$
\STATE $u_m^{[i+1]} = u_m^{[i]}-t_{2,m}^{[i]}+q_{2,m}^{[i]}$
\ENDFOR
\STATE \textsl{Update}
\STATE {$\begin{bmatrix}
y^{[i+1]}\\
\bh^{[i+1]}
\end{bmatrix}
= \begin{bmatrix}
y^{[i]}\\
\bh^{[i]}
\end{bmatrix}
-\gamma^{[i]}\left(\nabla\Psi
\Big(\Big[\begin{smallmatrix} s_{1}^{[i]}\\t_{1}^{[i]} \end{smallmatrix}\Big]\Big) + \left[\begin{smallmatrix}
F^* w_{1}^{[i]}\\
\sum_{m=1}^3 w_{2,m}^{[i]}
\end{smallmatrix}\right] \right)$ }
\ENDFOR
\end{algorithmic}
\end{algorithm}

\section{Results}
\label{sec:results}
\subsection{Evaluation methodology}\label{sec:results-methodology}
We consider either  synthetic or  real data for our evaluations. The first ones are evaluated both qualitatively and quantitatively, and the choice for the sparsity norm for the wavelet coefficients is discussed.
Realistic synthetic data are obtained from a modeled seismic trace with  primaries $y$.  
	Two multiple templates  ($J=2$)   $r_0$ and $r_1$
are independently convolved with  time-varying filters and summed up to yield a known, realistic, synthetic secondary reflection signal $s$. The primaries are  then corrupted by $s$ and an additive Gaussian noise.
The $j$-th time-varying filter is built upon  averaging filters with  length $P_j$, such that, $\forall p \in  \{p',\ldots,p'+P_j-1\}$, $\overline{h}_j^{(n)}(p) = \eta_j^{(n)}/P_j$ (cf. Eq. \eqref{e:sn}). 
The time-varying filters are thus unambiguously defined, at a given time $n$, by the constants $\eta_j^{(n)}$.
Uniform filters are chosen for their poor frequency selectivity behavior and notches in the frequency domain. Such artifacts for instance  happen in marine seismic acquisition. 
\subsection{Qualitative results on simulated data}\label{sec:simuls-results-lcd}

\begin{figure}[tbh]
\centering
\includegraphics[width=8.8cm,height=4.5cm]{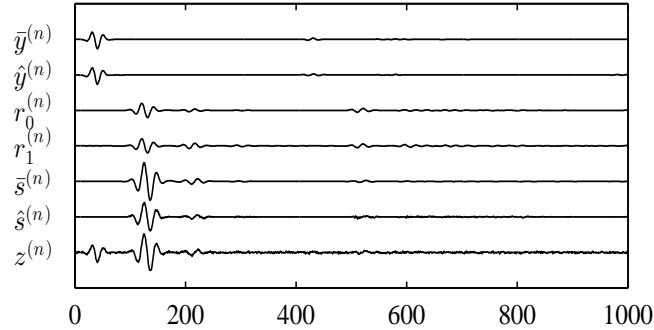}
\caption{Considered simulated seismic signals with noise level $\sigma=0.08$. From top to bottom: primary (unknown) $\bar{y}$, estimated $\hat{y}$, first template $r_0$, second template $r_1$, multiple (unknown) $\bar{s}$, estimated $\hat{s}$, and observed signal $z$. \label{fig:simuls_all_001}}
\end{figure}

\begin{figure}[tbh]
\centering
\includegraphics[width=8cm]{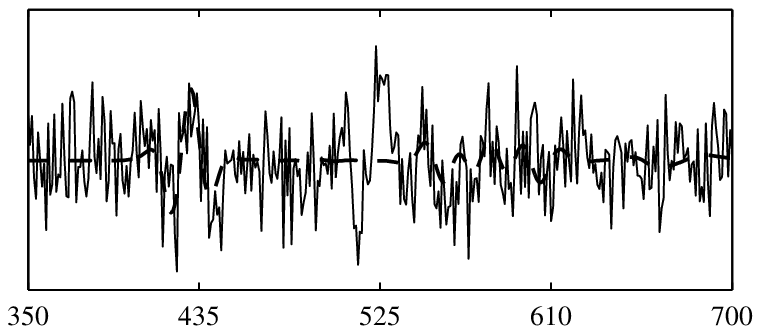}\hspace{0.05mm}
\includegraphics[width=8cm]{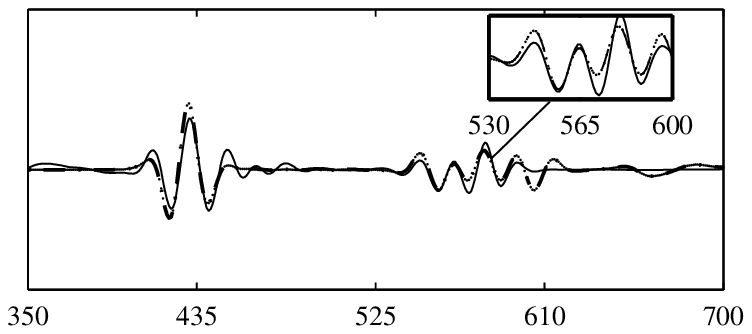}
\caption{Close-up with $\sigma=0.08$ and $\widetilde{\rho}$ is the $\ell_{1,2}$-norm; top: input data $z$ (solid), primary $\bar{y}$ (dashed);  bottom:  output separated primary $\hat{y}$ (dash-dotted) and primary $\bar{y}$ (solid).\label{fig:synthetic1D_008}}
\end{figure}

We choose here  two filter families with   lengths $P_0=10$ and $P_1=14$. 
The two filters evolve complementally in time, emulating a bi-modal multiple mixture at two different depths. 
They are combined with the two templates in a multiple signal depicted at the fifth row from the top of Fig.~\ref{fig:simuls_all_001}. Data and template were designed in order to mimic the time and frequency contents of seismic signals.
This figure also displays the other signals of interest, known and unknown, when $\sigma = 0.08$.
We aim at recovering weak primary signals, potentially hidden under both multiple and random perturbations, from the observed signal $z$ at the last row. We focus   on the rightmost part of the plots, between indices 350 and 700. The primary 
events we are interested in are located in Fig.~\ref{fig:simuls_all_001} 
(first signal on top) between indices 400-500 and 540-600, respectively. These primary events are mixed with multiples and random noise. A close-up on indices 350-700 is provided in Fig. \ref{fig:synthetic1D_008}. The first interesting primary event (400-500) is mainly affected by the random noise component. It serves as a witness for the quality of signal/random noise separation, as it is relatively insulated. The second one is disturbed by both noise and a multiple signal, relatively higher in amplitude than the first primary event. Consequently, its recovery severely probes the efficiency of the proposed algorithm.

\begin{figure}[htb]
\centering
\includegraphics[width=8.8cm]{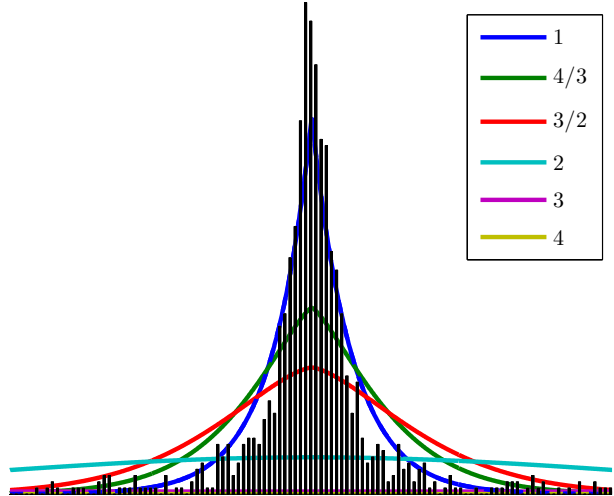}
\caption{Generalized Gaussian modeling of  seismic data wavelet  frame decomposition with different power laws.\label{fig_power_law}}
\end{figure}

For the proposed method, we choose the following initial settings. An undecimated wavelet frame transform with 8-length Symmlet filters is performed  on $\mathcal{L}=4$ resolution levels.  The loss functions $\psi$ and $(\widetilde{\varphi}_\ell)_{1\le \ell \le \mathcal{L}}$ in \eqref{eq:fct_constrain} are chosen as $\psi= \| \, . \,\|^2$ and $\widetilde{\varphi}_\ell=\,|\,.\,|$. 
The latter is based on a selection of power laws (namely, 1, 4/3, 3/2, 2, 3, and 4) for which closed-form proximal operators exist \cite[p. 1356]{Combettes_P_2007_j-siam-optim_proximal_tamob}. 
The best matching power for the chosen wavelet tight frame yields the taxicab metric or $\ell_1$-norm, as illustrated in Fig. \ref{fig_power_law}. 

The constraints $C_1$ and $C_2$ are chosen according to \eqref{eq:C_1} and \eqref{eq:C_2}, with  $\varepsilon_{1,p}= 0.1$ and $\varepsilon_{2,p}= 0.07$ for every $p$.  The bounds of the constraints are calculated empirically on ideal signals. For real signals, we propose to infer those constants from other methods. In practice, alternative cruder filtering or restoration algorithms indeed exist, with  the same purpose. They often are less involved and accurate, and potentially faster. They are run for instance on a small subset of representative real data. Thus, we obtain a first set of solutions, here separating primaries and multiples. Approximate constraints, required in the proposed method, are then computed (in a relatively fast manner) on approximate versions of unknown clean signals.
 Such a procedure yields coarse bound  estimates, upon which the proposed algorithm is run. Although approximate, they are expected to be easier to estimate than regularization hyper-parameters.
We hereafter use a fast first-pass separation of signals using \cite{Ventosa_S_2012_j-geophysics_adaptive_mswbcuwf}.
Finally, $\widetilde{\rho}$ is chosen as the $\ell_{1,2}$-norm.

\begin{figure}[tbh]
\centering
\includegraphics[width=8cm]{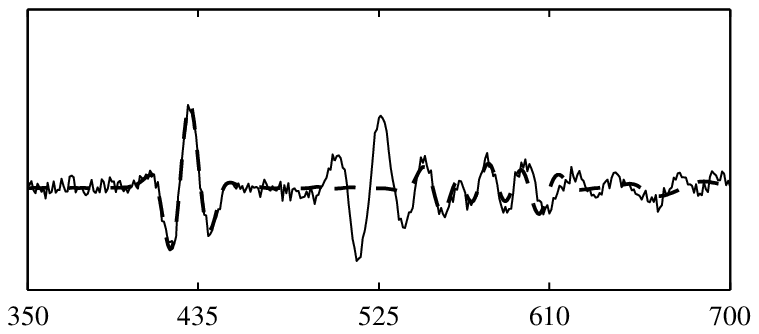} \hspace{0.05mm}
\includegraphics[width=8cm,]{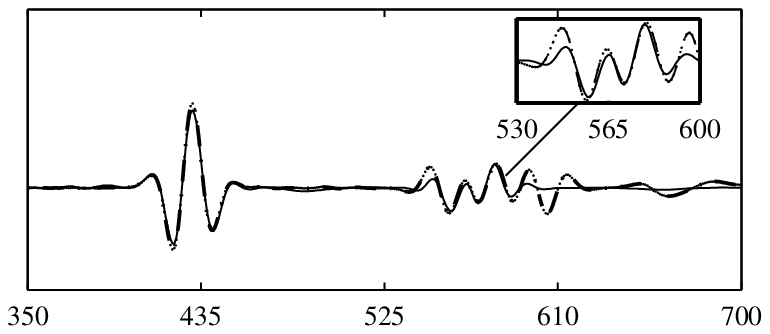}
\caption{Close-up with $\sigma=0.01$ and $\widetilde{\rho}$ is the $\ell_{1,2}$-norm; top: input data $z$ (solid), primary $\bar{y}$ (dashed);  bottom:  output separated primary $\hat{y}$ (dash-dotted) and primary $\bar{y}$ (solid).\label{fig:synthetic1D_001}}
\end{figure}

Fig.  \ref{fig:synthetic1D_001} and Fig.  \ref{fig:synthetic1D_008} provides close-ups of delimited areas with weak primaries. With a low noise level (Fig. \ref{fig:synthetic1D_001}), the multiple echo at indices 500 to 550 is faithfully removed, as well as the random noise. The first strong primary (indices 400-450) is well recovered. The second one has its first four periods matched correctly. More interesting is the stronger noise condition in Fig. \ref{fig:synthetic1D_008}. The multiple signal is still correctly removed and the incoherent noise is drastically reduced. Although with a noticeable amplitude distortion and some ringing effect, both primaries are visually recovered. As stated in Section \ref{sec:related-work}, the ability to restore --- albeit imperfectly --- spurs of strongly hidden primaries (cf.  Fig. \ref{fig:synthetic1D_008}, between indices 540 and 600) is of paramount interest for seismic  exploration in greater depths. 
\subsection{Quantitative results on simulated data}\label{sec:simuls-results-objective-lcd}
These first  qualitative simulation results are complemented with more extensive tests on different settings of  wavelet choices, levels, redundancy  and adaptive filter norms,  to limit the risk of unique parameter set  bias effects. We test three different dyadic wavelets (Haar, Daubechies and Symmlet with filter length 8), either in orthogonal basis or shift-invariant tight frame mode, with 3 or 4 decomposition levels, consistent with seismic data bandwidth. These data decomposition settings are tested again for four noise levels and three different choices of concentration metrics for the adaptive filters.   For each choice in  this parameter set, 100 different noisy realizations are processed. Each of these experiments is represented by its empirical average  and standard deviation. As we have seen before, the restoration of primaries or the cancellation of multiples could be jointly pursued.  We thus report in Tables~\ref{tab:snr_y} and \ref{tab:snr_s}  the average SNR, for the clean modeled primary and  multiple, with respect to their restored counterpart, respectively. Column headers $b$ and $f$ denote basis and frame results, whose averages are loosely denoted by $\mu_{\abbBasis}$ and $\mu_{\abbFrame}$. To improve reading, numbers in bold (for Table \ref{tab:snr_s} as well) indicate the best result for a given decomposition level  $\mathcal{L}$. Numbers in italics  denote the best SNRs obtained, irrespective of the number of wavelet levels. 
The standard deviation tables are combined with  Tables~\ref{tab:snr_y} and \ref{tab:snr_s} in Table \ref{tab_student_y_s} as a ``significance index'' of the outcome. 

  We first exemplify results in the  leftmost part of Table~\ref{tab:snr_y}  (Haar wavelet), for the first two rows. 
For the $\ell_1$-norm, we observe a primary restoration improvement of \SI{2.2}{\decibel} ($21.3-19.1$~dB, with standard deviations of \SI{0.16}{\decibel} for frames and \SI{0.26}{\decibel} for bases) for 3 wavelet levels. For 4 levels, we obtain \SI{1.2}{\decibel} with standard deviations of \SI{0.18}{\decibel} and \SI{0.22}{\decibel}. Intuitively, the SNR improvement appears to be significant, relatively to the dispersion. We shall detail this aspect later on. Yet,  such a sensible variation of about \SI{1}{\decibel}, with only one further wavelet decomposition level, further justifies the need for the given multi-parameter analysis. 

In most cases, for frames, the best results are obtained with 4 levels. When this is not the case, the difference in performance generally lies within the dispersion. This assertion cannot be stated with bases, possibly due to shift variance effects. 
The frame-based SNR for primaries is always greater, or equal to that of the basis one, putting statistical significance aside for the moment.  Namely, looking at summary statistics, the minimum, median, mean and  maximum improvements for frames over bases are \SI{0.5}{\decibel}, \SI{1.8}{\decibel}, \SI{2}{\decibel} and  \SI{4.2}{\decibel} respectively. Looking at numbers in bold, we see that a frame with the $\ell_2$ loss function is the clear winner in absolute SNR, for every wavelet choice and noise level.

The results for multiple estimation, given in Table \ref{tab:snr_s}, are more contrasted. Frames and bases yield more similar performance, especially  for high Gaussian noise levels. The best overall results (bold) are given by  $\ell_{1,2}$ (high noise) and  $\ell_{2}$-norms (low  noise). 

\begin{figure}[htb]
\centering
\includegraphics[width=8.8cm]{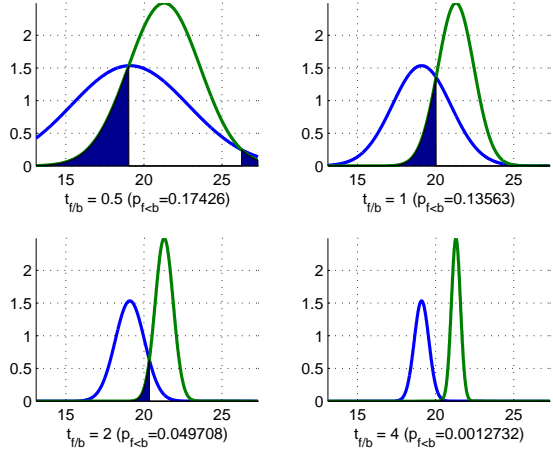}
\caption{Significance index abacus with different ``significance levels'' (and probability $\pi_{\abbFrameVsBasis}$, shaded): 0.5 (0.17), 1 (0.14), 2 (0.05) and 4 (0.0013).\label{significance}}
\end{figure}

Average differences  allow the observation of global trends. In practice, consistent results, taking into account SNR dispersion, are more important. Assuming denoised realizations follow a Gaussian distribution, we now study the difference Gaussian pdf between frame and basis results.  Its mean is $\mu_{\abbFrameVsBasis} = \mu_{\abbFrame}-\mu_{\abbBasis}$ and its variance $\sigma_{\abbFrameVsBasis}^2$ is  $\sigma_{\abbBasis}^2+\sigma^2_{\abbFrame}$. Table \ref{tab_student_y_s} reports the normalized difference significance  index $t_{\abbFrameVsBasis} = \mu_{\abbFrameVsBasis}/\sigma_{\abbFrameVsBasis}$, reminiscent of the Student's test. It is associated with the probability $\pi_{\abbFrameVsBasis}$ that, in  an outcome of the realizations, the basis SNR is superior to the frame SNR. 
The Haar wavelet,  $\ell_1$-norm,  primary restoration improvement of $\mu_{\abbFrameVsBasis} = \SI{2.2}{\decibel}$ yields $\sigma_{\abbFrameVsBasis} = 0.30$. Hence,  $t_{\abbFrameVsBasis}  = 7.2$. The interpretation of this significance is illustrated with the abacus in Figure \ref{significance}, with $\pi_{\abbFrameVsBasis}$ associated to the shaded area.
We deem the difference in distribution between bases and frames significant only if $|t_{\abbFrameVsBasis} |> 1$. When significant, the index is emphasized in italics or bold, the latter denoting the most significant among the three concentration measures.
Since  the minimum, median, mean and  maximum indices  $t_{\abbFrameVsBasis}$ are 1.3, 3.4, 3.5 and 7.7, we consider the  improvement of frames over bases significant for all the tested parameters for primaries  (Table \ref{tab_student_y_s}-left). Interestingly, whereas the $\ell_2$-norm gives the best average gain, the $\ell_{1,2}$ and the $\ell_{1}$-norms yield sensibly more significance at lower noises. Choosing the $\ell_{1,2}$  or $\ell_{1}$-norm is consequently more interesting in practice, as we desire more consistent results under unknown noise variations in observed signals. At higher noise levels, the normalized difference index  $t_{\abbFrameVsBasis}$ (between values 3 and 4) is very close for all parameters and wavelets. Thus, the significance of the different filter concentration measures is not fundamentally different. Based on the previous observations, the choice of $\ell_{1,2}$ or $\ell_{1}$-norms would yield more consistent results in all cases.

The rightmost and lower part of Table	\ref{tab_student_y_s} concerns the restoration of multiples. Although of weaker importance in practice, we notice that   most  $t_{\abbFrameVsBasis}$ values vary  between -1 and 1. This indicates that bases and frames perform similarly, since there is no significant   performance difference. This phenomenon  can be explained  by the fact that frames or bases, through the wavelet transform sparsity prior,  impact  primaries rather than  multiples.  
In few cases though, we observe, at low noise levels, some significance in frame outcomes over basis results, obtained by  the $\ell_{1,2}$ and  $\ell_{1}$-norms again.

Globally, the restoration of both  primaries and multiples benefits from the choice of  a Daubechies or Symmlet wavelet frame. The best performance --- in terms of statistical significance --- is offered by sparsity-promoting $\ell_{1}$ and  $\ell_{1,2}$-norms, either at  lower and  higher noise level.

\begin{table}[htb]\centering\begin{tabular}{|c|c|c|c|c|c|c|c|c|c|c|c|c|c|c|c|c|c|c|c|}
\cline{3-20}
\multicolumn{2}{c}{} & \multicolumn{18}{|c|}{ mean$(\text{SNR}_y)$} \\ 
\cline{3-20}
\multicolumn{2}{c}{} & \multicolumn{6}{|c}{ Haar }  & \multicolumn{6}{|c}{ Daubechies }  & \multicolumn{6}{|c|}{ Symmlet } \\ 
\cline{1-20}
\multicolumn{2}{|c}{$\widetilde{\rho}$} & \multicolumn{2}{|c}{$\ell_1$} & \multicolumn{2}{|c}{$\ell_2$} & \multicolumn{2}{|c}{$\ell_{1,2}$} & \multicolumn{2}{|c}{$\ell_1$} & \multicolumn{2}{|c}{$\ell_2$} & \multicolumn{2}{|c}{$\ell_{1,2}$} & \multicolumn{2}{|c}{$\ell_1$} & \multicolumn{2}{|c}{$\ell_2$} & \multicolumn{2}{|c|}{$\ell_{1,2}$}  \\
\hline
$\sigma$ & $\mathcal{L}$ & \abbBasis & \abbFrame & \abbBasis & \abbFrame & \abbBasis & \abbFrame & \abbBasis & \abbFrame & \abbBasis & \abbFrame & \abbBasis & \abbFrame & \abbBasis & \abbFrame & \abbBasis & \abbFrame & \abbBasis & \abbFrame \\ 
\hline
0.01 & 3 & 19.1 &  21.3 & 20.2 &  \tabEmphA{21.9} &  18.9 & 21.3& \tabEmphB{20.9} &  21.8 &  21.1 & \tabEmphA{22.3} &  20.1 & 21.8 & 19.9 &  21.8 &21.4&  \tabEmphA{22.5} &  19.2 & 21.9 \\ 
\cline{2-20}
& 4 &  \tabEmphB{20.6} & \tabEmphB{21.8}& \tabEmphB{21.6}  & \tabEmphA{\tabEmphB{22.4}}& \tabEmphB{20.3} & \tabEmphB{21.8}&  20.8 & \tabEmphB{22.0}&\tabEmphB{22.1} & \tabEmphA{\tabEmphB{22.7}} &  \tabEmphB{21.1} & \tabEmphB{22.1} &  \tabEmphB{20.3} & \tabEmphB{22.0}&   \tabEmphB{22.3} & \tabEmphA{\tabEmphB{22.8}}&  \tabEmphB{20.4} & \tabEmphB{22.1}\\ 
\hline
0.02 & 3 &  19.2 &  20.3 & 20.1  &  \tabEmphA{21.1} & 18.6 &  20.1 &  \tabEmphB{20.6} &  21.3 & \tabEmphB{21.2}&  \tabEmphA{22.1} &  \tabEmphB{20.4} &  21.2 &  \tabEmphB{20.1} &  21.5 & 21.4&  \tabEmphA{22.3} &  \tabEmphB{19.5} &  21.3 \\ 
\cline{2-20}
& 4 &  \tabEmphB{19.8} & \tabEmphB{20.8}& \tabEmphB{20.5} & \tabEmphA{\tabEmphB{21.3}}& \tabEmphB{19.5} & \tabEmphB{20.6} &  20.1 & \tabEmphB{21.6}&  21.0 & \tabEmphA{\tabEmphB{22.2}}&  20.0 & \tabEmphB{21.5} &  20.0 & \tabEmphB{21.6}&  \tabEmphB{21.5} & \tabEmphA{\tabEmphB{22.4}}&  19.2 & \tabEmphB{21.6}\\ 
\hline
0.04 & 3 & 16.5 &   18.2 & 16.9  & \tabEmphA{18.5}&  16.2 & 18.0 & \tabEmphB{18.0} & 20.1 & 18.3& \tabEmphA{20.6} &  \tabEmphB{17.9} & 19.9 & \tabEmphB{17.9} & 20.2& \tabEmphB{18.4}&  \tabEmphA{\tabEmphB{20.8}} &  \tabEmphB{17.6} & 20.0\\ 
\cline{2-20}
& 4 &  \tabEmphB{16.7} &  \tabEmphB{18.5} &  \tabEmphB{16.9} & \tabEmphA{\tabEmphB{18.8}}& \tabEmphB{16.5} &  \tabEmphB{18.3} & 18.0 &  \tabEmphB{20.2} &  \tabEmphB{18.3} & \tabEmphA{\tabEmphB{20.7}}&  17.7 & \tabEmphB{20.0}  &  17.9 &  \tabEmphB{20.3} &  18.21 &\tabEmphA{20.7}&  17.5 &  \tabEmphB{20.1} \\ 
\hline
0.08& 3 &  12.3 &  14.7 & 12.5   &\tabEmphA{\tabEmphA{14.9}} &  12.2 & 14.6  &  \tabEmphB{14.3} &  17.4 & \tabEmphB{14.4}& \tabEmphA{17.7}& \tabEmphB{14.2}& 17.3  &  \tabEmphB{14.1} & \tabEmphB{17.6}& \tabEmphB{14.3}& \tabEmphA{\tabEmphB{17.9}}& \tabEmphB{14.0}& \tabEmphB{17.5}\\
\cline{2-20}
& 4 &  \tabEmphB{12.4} & \tabEmphB{15.0} & \tabEmphB{12.5} & \tabEmphA{\tabEmphB{15.2}} &  \tabEmphB{12.3} & \tabEmphB{14.9} &  14.2 & \tabEmphB{17.5} &  14.4 & \tabEmphA{\tabEmphB{17.7}} &  14.1 & \tabEmphB{17.3} &  13.9 & 17.5 &  14.0 &  \tabEmphA{17.7} &  13.2 &  17.3 \\ 
\hline
\end{tabular}\caption{SNR, averaged over $100$ noise realizations for the estimations of $y$. \label{tab:snr_y}}\end{table}

\begin{table}[htbp]\centering\begin{tabular}{|c|c|c|c|c|c|c|c|c|c|c|c|c|c|c|c|c|c|c|c|}
\cline{3-20}
\multicolumn{2}{c}{} & \multicolumn{18}{|c|}{ mean$(\text{SNR}_s)$} \\ 
\cline{3-20}
\multicolumn{2}{c}{} & \multicolumn{6}{|c}{ Haar }  & \multicolumn{6}{|c}{ Daubechies }  & \multicolumn{6}{|c|}{ Symmlet } \\ 
\cline{1-20}
\multicolumn{2}{|c}{$\widetilde{\rho}$} & \multicolumn{2}{|c}{$\ell_1$} & \multicolumn{2}{|c}{$\ell_2$} & \multicolumn{2}{|c}{$\ell_{1,2}$} & \multicolumn{2}{|c}{$\ell_1$} & \multicolumn{2}{|c}{$\ell_2$} & \multicolumn{2}{|c}{$\ell_{1,2}$} & \multicolumn{2}{|c}{$\ell_1$} & \multicolumn{2}{|c}{$\ell_2$} & \multicolumn{2}{|c|}{$\ell_{1,2}$}  \\
\hline
$\sigma$ & $\mathcal{L}$ & \abbBasis & \abbFrame & \abbBasis & \abbFrame & \abbBasis & \abbFrame & \abbBasis & \abbFrame & \abbBasis & \abbFrame & \abbBasis & \abbFrame & \abbBasis & \abbFrame & \abbBasis & \abbFrame & \abbBasis & \abbFrame \\ 
\hline
0.01 & 3 &  26.3 &  27.0 & 27.4   & \tabEmphA{28.4} &  26.2 &  28.0 & \tabEmphB{27.6} &  28.0 &  28.2 & \tabEmphA{28.3} &  26.9 &  28.0 & 26.7 &  28.0 &  27.9 &  \tabEmphA{28.4} & 26.5& 28.0\\ 
\cline{2-20}
& 4 & \tabEmphB{27.6} & \tabEmphB{28.3}& \tabEmphB{28.4}  & \tabEmphA{\tabEmphB{28.7}}&  \tabEmphB{27.4} & \tabEmphB{28.4} &  27.6 & \tabEmphB{28.1}&  \tabEmphA{\tabEmphB{28.6}} &  \tabEmphB{28.5} &  \tabEmphB{27.7} & \tabEmphB{28.1} &  \tabEmphB{27.1} & \tabEmphB{28.1}&  \tabEmphB{28.5} & \tabEmphA{\tabEmphB{28.6}}&  \tabEmphB{27.1} & \tabEmphB{28.2}\\ 
\hline
0.02 & 3 &  25.3 &  25.6 &\tabEmphA{25.7}  & \tabEmphA{25.7} &  25.0 & 25.5  & \tabEmphA{\tabEmphB{25.8}} &  25.5 & \tabEmphB{25.8} & 25.6 & \tabEmphA{\tabEmphB{25.8}} &  25.5 & \tabEmphB{25.3} &  25.5 & \tabEmphA{25.8} &  25.6 &  25.1 &  25.5 \\ 
\cline{2-20}
& 4 & \tabEmphB{25.8} & \tabEmphB{25.7 }&  \tabEmphA{\tabEmphB{26.0}} & \tabEmphB{25.8 } &  \tabEmphB{25.7} & \tabEmphB{25.8} &  25.5 & \tabEmphB{25.6 }&  25.7 & \tabEmphA{\tabEmphB{25.6}} &  25.5 & \tabEmphA{\tabEmphB{25.6}}&  25.3 & \tabEmphB{25.6}&  \tabEmphA{\tabEmphB{25.9}} & \tabEmphB{25.6}&  \tabEmphB{25.3} & \tabEmphB{25.6}\\ 
\hline
0.04 & 3 &  22.1 &  22.1 & 21.8  &  21.8 &  22.2 &  \tabEmphA{22.3} &  \tabEmphB{22.3} &  22.1 &  20.1 &  \tabEmphB{21.8} &  \tabEmphA{\tabEmphB{22.5}} & \tabEmphB{22.3} &  \tabEmphB{22.1} &  22.1 &  21.8 &  21.8 &  \tabEmphB{22.2} &  \tabEmphA{\tabEmphB{22.3}}\\ 
\cline{2-20}
& 4 &  \tabEmphB{22.3} &  \tabEmphB{22.2} &  \tabEmphB{21.9} & \tabEmphB{21.8}&  \tabEmphA{\tabEmphB{22.4}} & \tabEmphB{22.3}&  22.2 &  \tabEmphB{22.1} & \tabEmphB{21.8} & \tabEmphB{21.8}&  \tabEmphA{22.4} & 22.2 &  21.2 &  \tabEmphB{22.1} &  \tabEmphB{21.8} &\tabEmphB{21.8}&  \tabEmphB{22.2} & \tabEmphA{22.3}\\ 
\hline
0.08 & 3 &  18.4 &  18.4 &  17.7  & \tabEmphB{17.8} & 18.6 & \tabEmphA{\tabEmphB{18.6}}&  \tabEmphB{18.5} &  18.4 &  \tabEmphB{17.8} & \tabEmphB{17.8} & \tabEmphA{\tabEmphB{18.7}} & \tabEmphB{18.6} &  18.4 &  18.4 &  \tabEmphB{17.8} &  \tabEmphB{17.8} & \tabEmphA{\tabEmphB{18.6}} & \tabEmphA{\tabEmphB{18.6}}\\ 
\cline{2-20}
& 4 & \tabEmphB{18.4} & \tabEmphB{18.4}& \tabEmphB{17.8}   &  \tabEmphB{17.8} & \tabEmphA{\tabEmphB{18.6}} & \tabEmphB{18.6}&  18.5 & \tabEmphB{18.4 }&  \tabEmphB{17.8} &  17.7 &  \tabEmphA{18.7} & 18.6 &  \tabEmphB{18.4} & \tabEmphB{18.4}&  \tabEmphB{17.8} &  17.8 &  \tabEmphA{18.6} &  18.6  \\ 
\hline
\end{tabular}\caption{SNR, averaged over $100$ noise realizations for the estimations of $s$. \label{tab:snr_s}}\end{table}

\begin{table}[htb]\centering\begin{tabular}{|c|c|c|c|c|c|c|c|c|c|c|c|c|c|c|c|c|c|c|c|}
\cline{3-20}
\multicolumn{2}{c}{} & \multicolumn{9}{|c}{ $t_{\abbFrameVsBasis}$ for primaries}  & \multicolumn{9}{|c|}{ $t_{\abbFrameVsBasis}$ for multiples} \\ 
\cline{3-20}
\multicolumn{2}{c}{} & \multicolumn{3}{|c}{ Haar }  & \multicolumn{3}{|c}{ Daubechies }  & \multicolumn{3}{|c}{ Symmlet }  & \multicolumn{3}{|c}{ Haar }  & \multicolumn{3}{|c}{ Daubechies }  & \multicolumn{3}{|c|}{ Symmlet } \\ 
\cline{1-20}
\multicolumn{2}{|c}{$\widetilde{\rho}$} & \multicolumn{1}{|c}{$\ell_1$} & \multicolumn{1}{|c}{$\ell_2$} & \multicolumn{1}{|c}{$\ell_{1,2}$} & \multicolumn{1}{|c}{$\ell_1$} & \multicolumn{1}{|c}{$\ell_2$} & \multicolumn{1}{|c}{$\ell_{1,2}$} & \multicolumn{1}{|c}{$\ell_1$} & \multicolumn{1}{|c}{$\ell_2$} & \multicolumn{1}{|c}{$\ell_{1,2}$}  &  \multicolumn{1}{|c}{$\ell_1$} & \multicolumn{1}{|c}{$\ell_2$} & \multicolumn{1}{|c}{$\ell_{1,2}$} & \multicolumn{1}{|c}{$\ell_1$} & \multicolumn{1}{|c}{$\ell_2$} & \multicolumn{1}{|c}{$\ell_{1,2}$} & \multicolumn{1}{|c}{$\ell_1$} & \multicolumn{1}{|c}{$\ell_2$} & \multicolumn{1}{|c|}{$\ell_{1,2}$} \\
\hline
$\sigma$ & $\mathcal{L}$ & \abbFrameVsBasis & \abbFrameVsBasis & \abbFrameVsBasis & \abbFrameVsBasis & \abbFrameVsBasis & \abbFrameVsBasis & \abbFrameVsBasis & \abbFrameVsBasis & \abbFrameVsBasis & \abbFrameVsBasis & \abbFrameVsBasis & \abbFrameVsBasis & \abbFrameVsBasis & \abbFrameVsBasis & \abbFrameVsBasis & \abbFrameVsBasis & \abbFrameVsBasis & \abbFrameVsBasis \\ 
\hline
0.01 & 3 & \tabEmphC{7.3} & \tabEmphC{4.3} & \tabEmphA{7.7} & \tabEmphC{2.6} & \tabEmphC{3.0} & \tabEmphA{4.9} & \tabEmphC{5.6} & \tabEmphC{2.6} & \tabEmphA{7.2} &\tabEmphC{2.7} & \tabEmphC{3.2} & \tabEmphA{7.5} & \tabEmphC{1.3} & 0.3 & \tabEmphA{3.5} & \tabEmphC{5.3} & \tabEmphC{1.5} & \tabEmphA{6.0} \\
\cline{2-20}
& 4 &  \tabEmphC{4.2} & \tabEmphC{2.8} & \tabEmphA{5.1} & \tabEmphA{4.2} & \tabEmphC{1.9} & \tabEmphC{3.3} & \tabEmphA{5.3} & \tabEmphC{1.3} & \tabEmphC{5.1}&\tabEmphC{2.9} & \tabEmphC{1.1} & \tabEmphA{4.2} & \tabEmphA{2.5} & -0.2 & \tabEmphC{1.6} & \tabEmphA{4.4} & 0.1 & \tabEmphA{4.4} \\
\hline
0.02 & 3 &  \tabEmphC{3.0} & \tabEmphC{2.7} & \tabEmphA{3.4} & \tabEmphC{1.5} & \tabEmphA{2.1} & \tabEmphC{1.6} & \tabEmphC{2.8} & \tabEmphC{2.1} & \tabEmphA{3.1}&0.7 & 0.0 & \tabEmphC{1.6} & \tabEmphC{-1.1} & -0.8 & -0.9 & 0.5 & -0.5 & \tabEmphA{1.5} \\
\cline{2-20}
& 4 &  \tabEmphA{2.6} & \tabEmphC{2.3} & \tabEmphC{2.5} & \tabEmphA{3.0} & \tabEmphC{2.9} & \tabEmphC{2.7} & \tabEmphC{3.0} & \tabEmphC{1.7} & \tabEmphA{4.3}&-0.1 & -0.6 & 0.1 & 0.3 & -0.3 & 0.2 & 0.8 & -0.7 & \tabEmphA{1.2} \\
\hline
0.04 & 3 & \tabEmphC{3.4} & \tabEmphA{3.5} & \tabEmphC{3.2} & \tabEmphC{3.0} & \tabEmphA{3.9} & \tabEmphC{2.7} & \tabEmphC{3.2} & \tabEmphA{3.8} & \tabEmphC{3.2}&0.0 & 0.0 & 0.1 & -0.6 & \tabEmphA{4.7} & -0.6 & -0.2 & -0.1 & 0.1 \\
\cline{2-20}
& 4 &  \tabEmphC{3.5} & \tabEmphA{3.7} & \tabEmphC{3.3} & \tabEmphC{3.2} & \tabEmphA{3.8} & \tabEmphC{2.8} & \tabEmphC{3.3} & \tabEmphA{3.7} & \tabEmphC{3.3}&-0.3 & -0.1 & -0.2 & -0.3 & -0.1 & -0.3 & \tabEmphA{2.6} & -0.1 & 0.1 \\
\hline
0.08 & 3 & \tabEmphA{3.5} & \tabEmphA{3.5} & \tabEmphA{3.5} & \tabEmphC{3.5} & \tabEmphA{3.9} & \tabEmphC{3.3} & \tabEmphC{3.8} & \tabEmphA{4.2} & \tabEmphC{3.7}&0.0 & 0.0 & 0.1 & -0.2 & 0.0 & -0.1 & -0.1 & 0.0 & 0.0 \\
\cline{2-20}
& 4 &  \tabEmphA{3.8} & \tabEmphA{3.8} & \tabEmphC{3.7} & \tabEmphC{3.4} & \tabEmphA{3.6} & \tabEmphC{3.2} & \tabEmphC{3.8} & \tabEmphC{4.1} & \tabEmphA{4.2}&0.0 & 0.0 & 0.0 & -0.1 & 0.0 & -0.2 & -0.1 & 0.0 & 0.0 \\
\hline
\end{tabular}\caption{Normalized difference significance  index.	\label{tab_student_y_s}}\end{table}

\subsection{Comparative evaluation: synthetic data\label{sec:synthetic-data-results}}
\begin{figure*}[tbh]
\centering
\subfigure[]{\includegraphics[width=8cm, height=5cm]{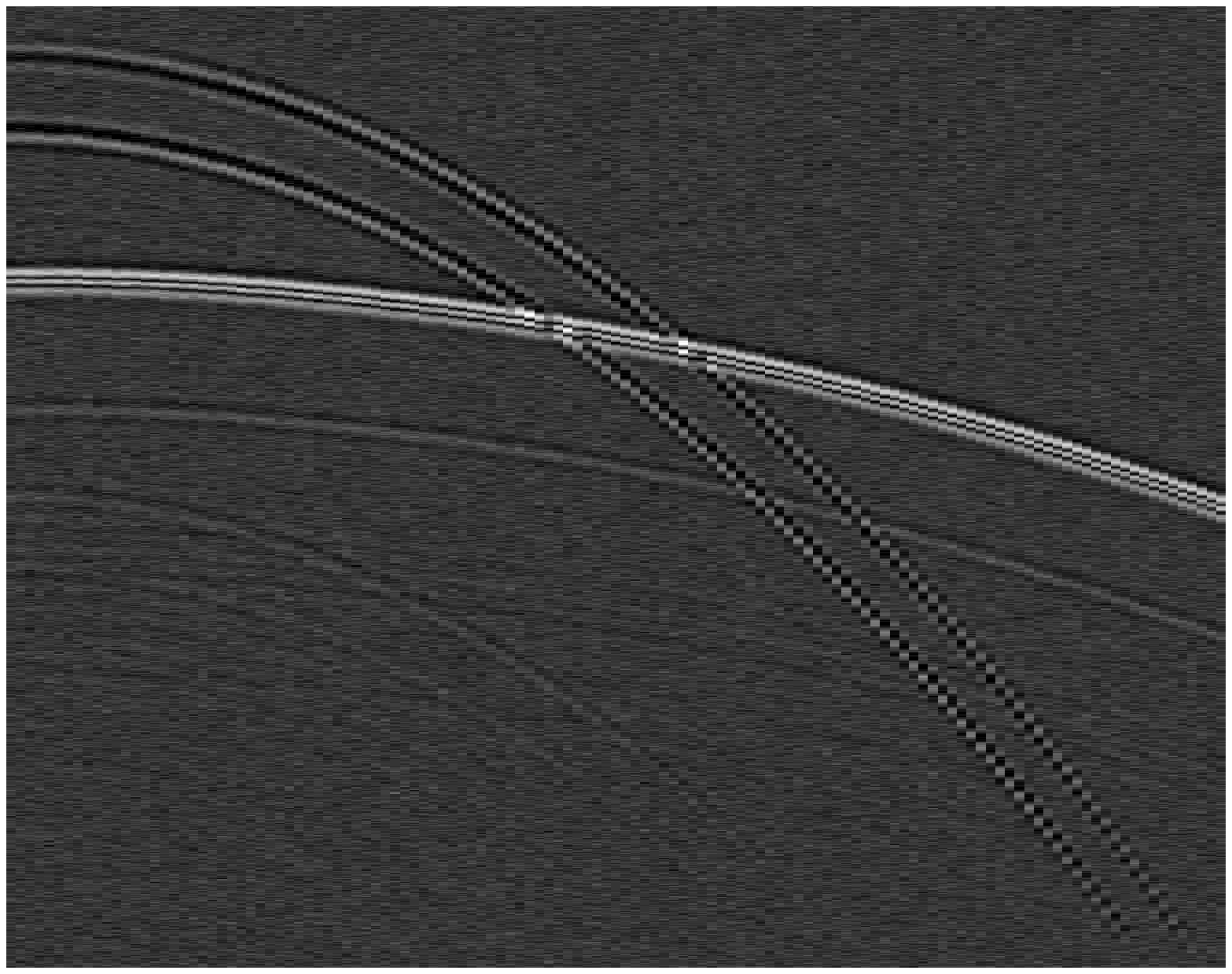}} \hspace{0.05mm}
\subfigure[]{\includegraphics[width=8cm, height=5cm]{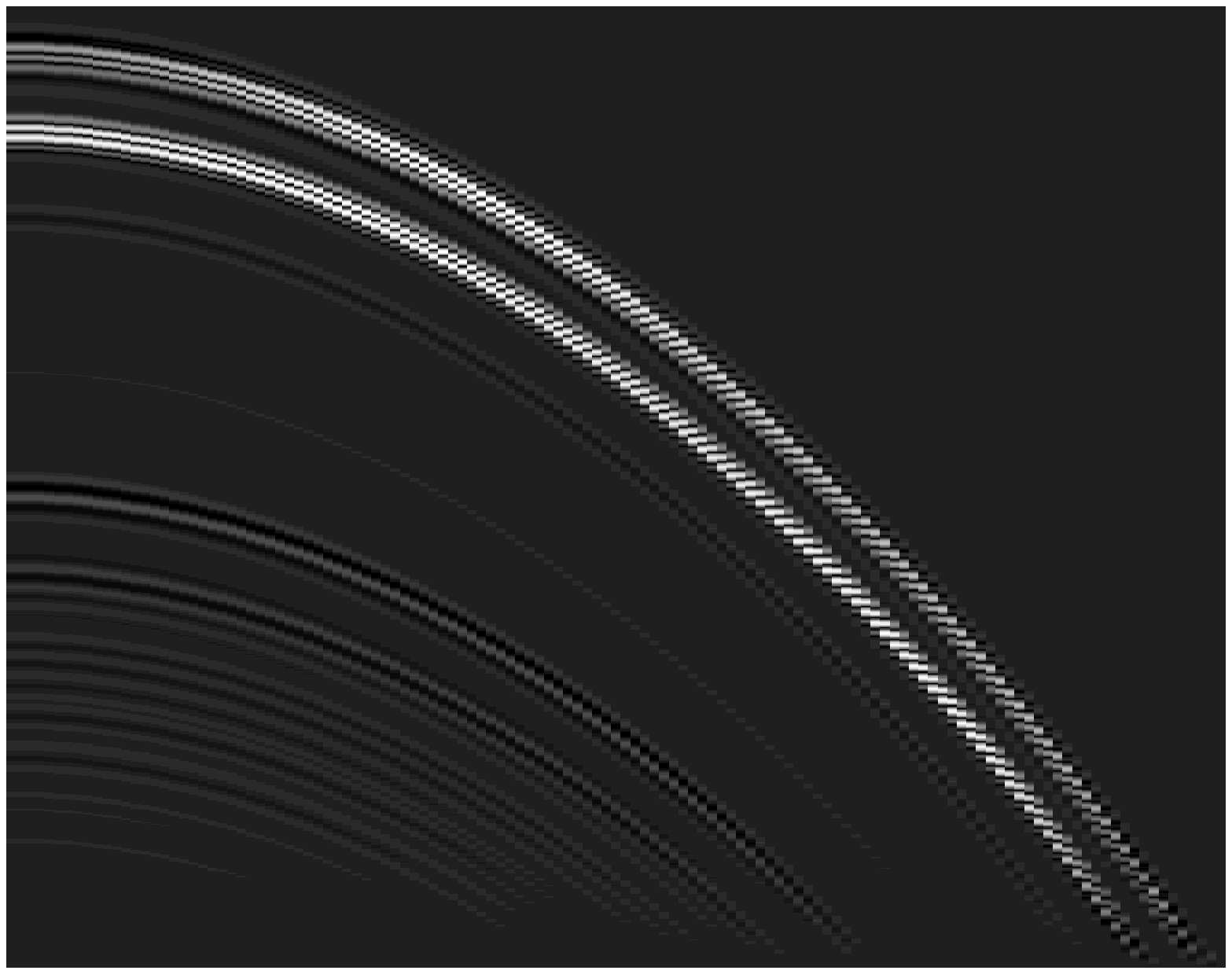}}\\
\subfigure[]{\includegraphics[width=8cm, height=5cm]{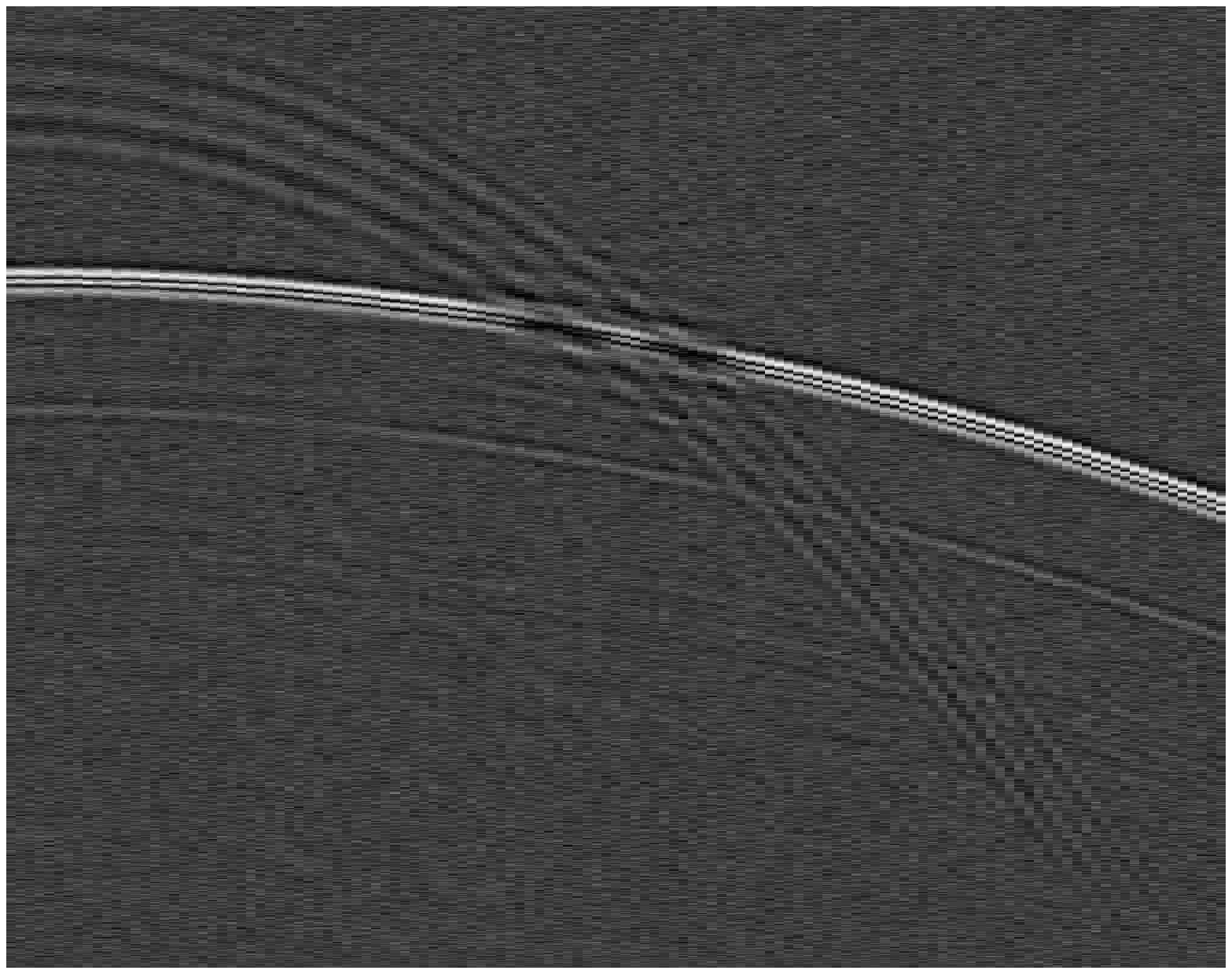}} \hspace{0.05mm}
\subfigure[]{\includegraphics[width=8cm, height=5cm]{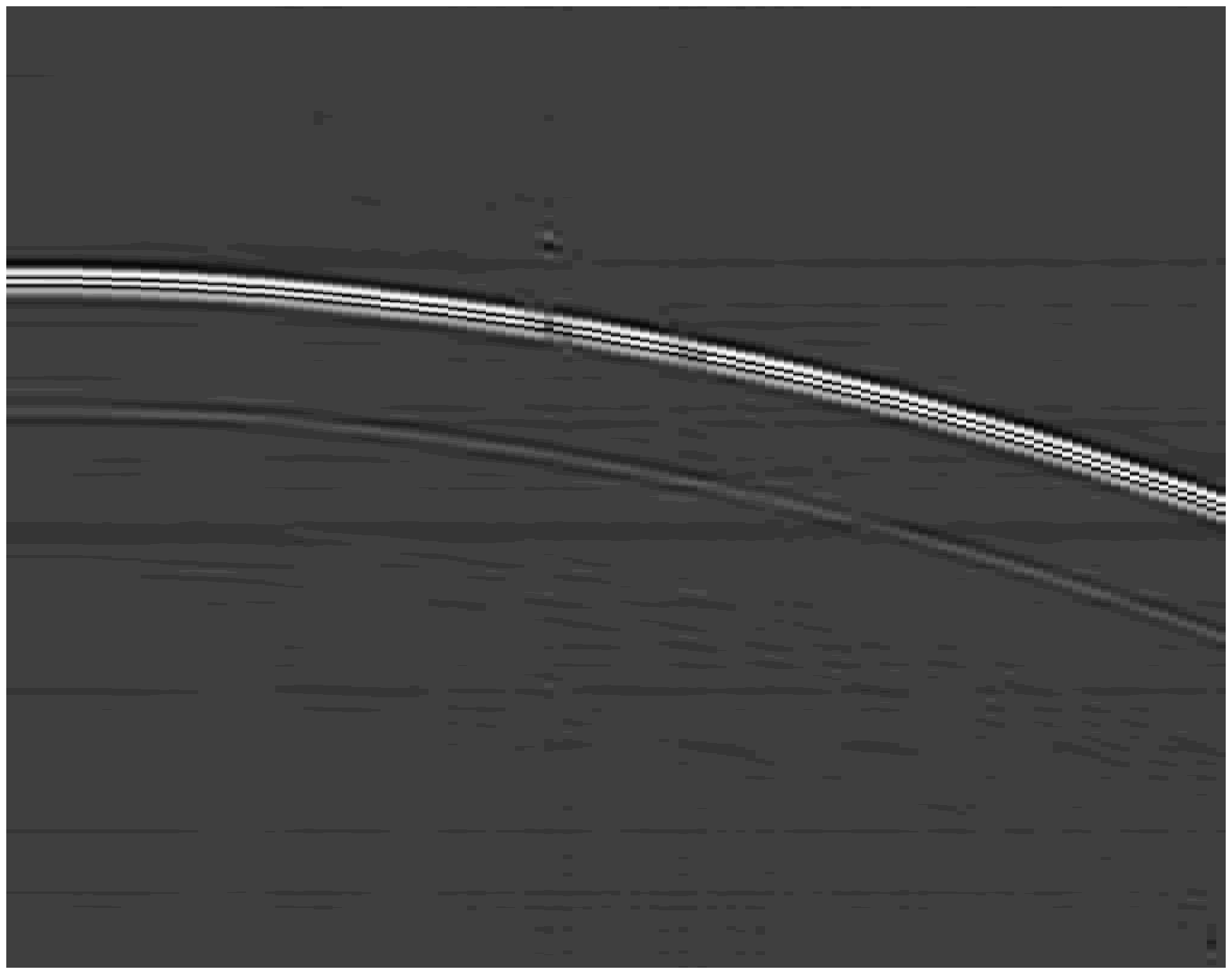}}\\
\subfigure[]{\includegraphics[width=8cm, height=5cm]{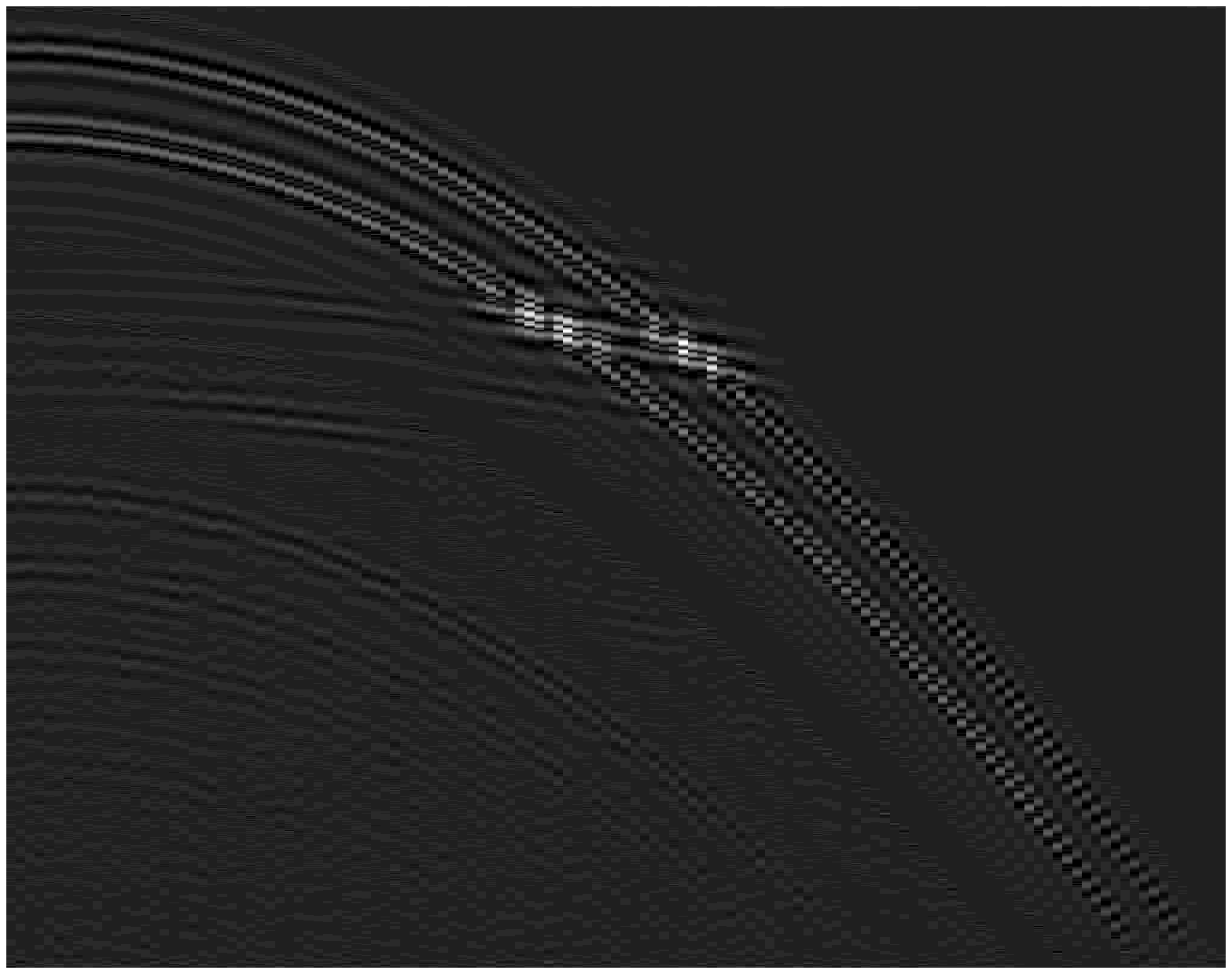}} \hspace{0.05mm}
\subfigure[]{\includegraphics[width=8cm, height=5cm]{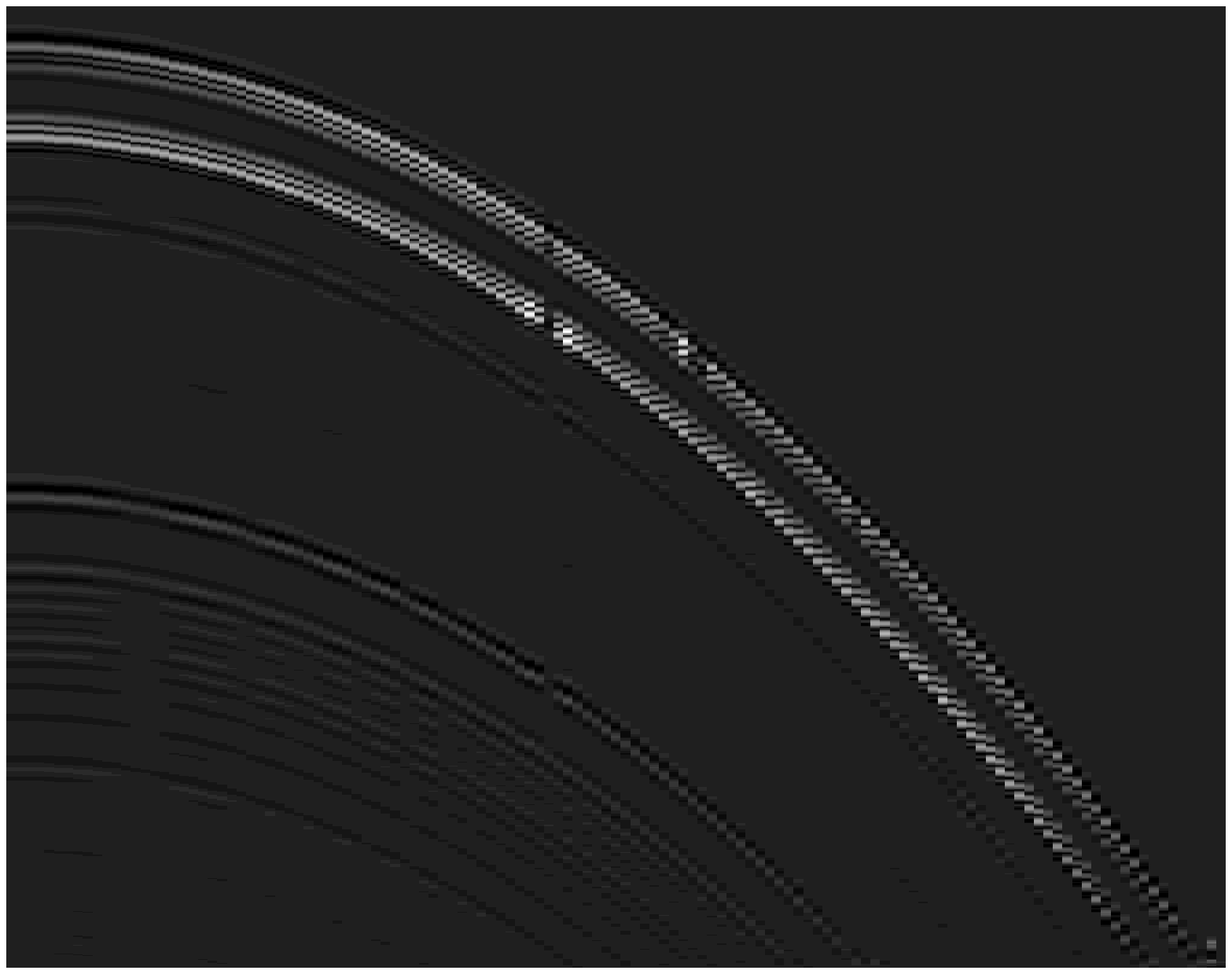}}
\caption{Data composed by three events ($\sigma = 0.08$), one primary and two multiples, $\textrm{SNR} =  \db{1.71}$ (a); multiples composed by two estimated events (b); output separated primaries with \cite{Ventosa_S_2012_j-geophysics_adaptive_mswbcuwf}, $\textrm{SNR} = \db{3.11}$ (c) and our method, $\textrm{SNR} = \db{16.77}$ (d); output adapted multiples with  \cite{Ventosa_S_2012_j-geophysics_adaptive_mswbcuwf}, $\textrm{SNR} = \db{3.1}$ (e) and our method, $\textrm{SNR} = \db{15.44}$ (f).\label{fig:synthetic2D_008}}
\end{figure*}

In addition to the above objective and subjective results, we perform a comparative evaluation with the empirical algorithm proposed in \cite{Ventosa_S_2012_j-geophysics_adaptive_mswbcuwf}. It is based on adaptive filtering on sliding windows in a complex continuous wavelet domain. The chosen complex Morlet wavelet is very efficient at concentrating seismic data energy. One-tap Wiener-like  (unary) complex filters are adaptively estimated in overlapping windows taken in the complex scalogram, i.e.  the complex-valued, discretized, continuous  wavelet transform. This algorithm was successfully tested against industry standards. It is quantitatively faster, but it does not permit the introduction of prior knowledge on signal sparsity or filter regularity. Fig. \ref{fig:synthetic2D_008} 
presents synthetic 2D seismic data, constructed similarly to the previous 1D traces in Section VI-B, with a high noise level ($\sigma = 0.08$). Vertical traces are stacked laterally to form a 2D image. From left to right,  the synthetic traces drift away from the seismic source.  The bended hyperbolas correspond to primaries. The flatter one, below, mimics a multiple event. 
Here, $P_0= 6$, $P_1 = 6$, and  constraints $C_1$ and $C_2$ are chosen according to (14) and (15), where  $\varepsilon_{1,p}= 0.1$ and $\varepsilon_{2,p}= 0.1$ for every $p$.
Apparently, better primary preservation is obtained with the proposed method, for a very simple synthetic data set. This phenomenon is observed at the crossing between primaries and multiples. The proposed method also effectively gets rid of more incoherent noise.

\subsection{Comparative evaluation: real data\label{sec:real-data-results}}

\begin{figure}[htb]
\centering
\subfigure[]{\includegraphics[width=8.8cm]{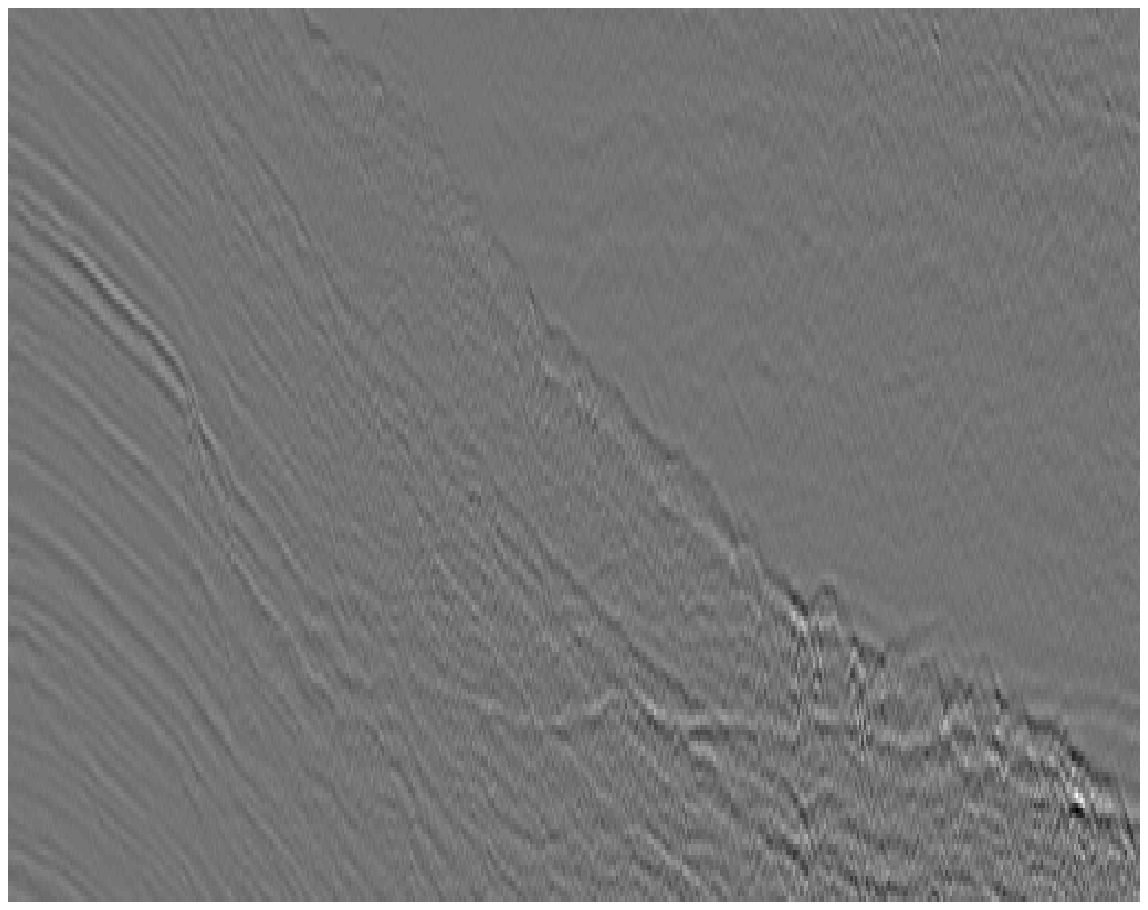}} \hspace{0.05mm}
\subfigure[]{\includegraphics[width=8.8cm]{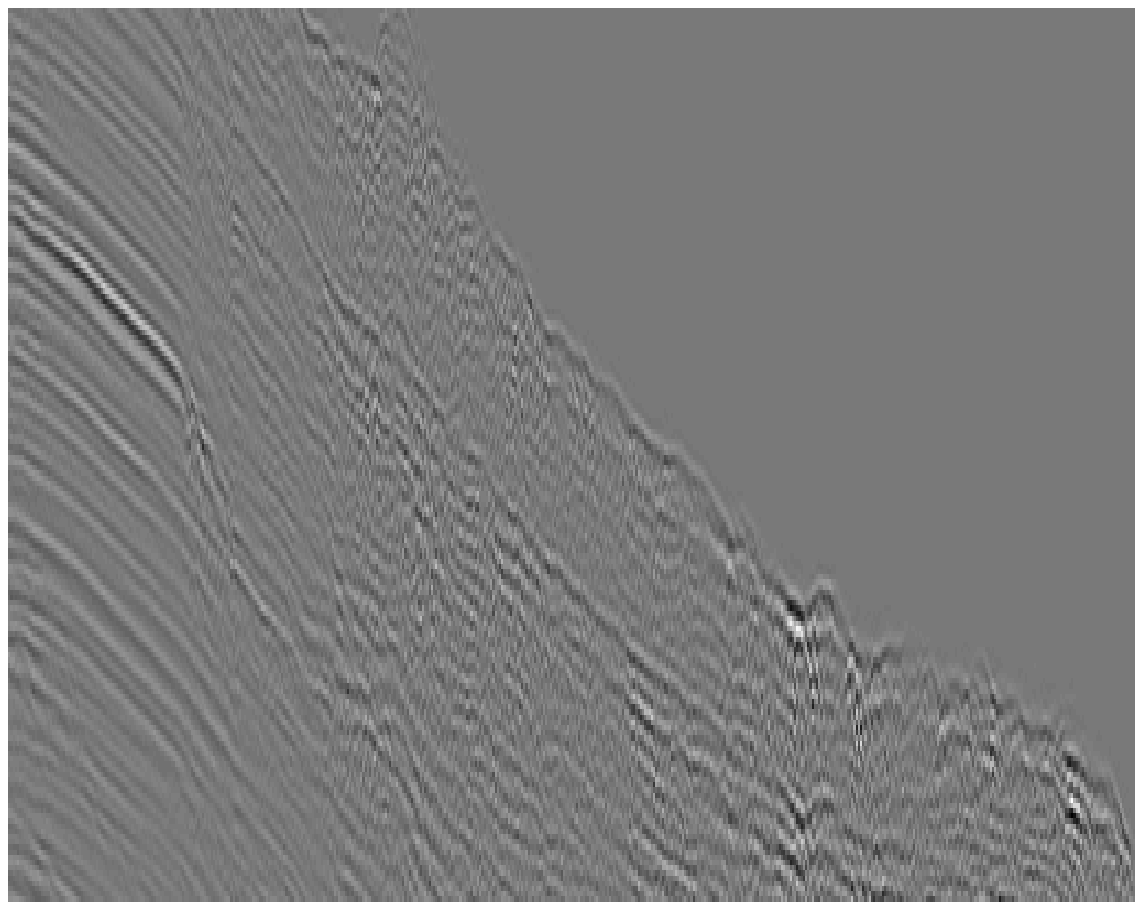}}\\
\caption{Portion of a common receiver  gather: (a) recorded seismic data with a partially appearing primary (b) multiple wavefield template.\label{fig:real2D_signal}}
\end{figure}

The previous simulated example is a little bit simplistic. We finally compare our algorithm with \cite{Ventosa_S_2012_j-geophysics_adaptive_mswbcuwf}  on a portion of a real seismic data set. Recorded  and multiple template data belong to the same marine seismic survey processed in \cite{Ventosa_S_2012_j-geophysics_adaptive_mswbcuwf}. The recorded seismic data is displayed in Fig. \ref{fig:real2D_signal}-(a). The main objective is to uncover a potential primary,  masked by strong multiple events that mostly contribute to the observed total seismic signal energy. The primary  appears partially as a wiggling, horizontal stripes in the bottom part of the figure, on the right side. Geologically speaking, it should be re-linked with the left side of the picture, to one of the dimmed sloping stripes.
By looking at differences between the recorded data and the multiple template in Fig. \ref{fig:real2D_signal}-(b), the trace of the flat primary may appear more obvious. Templates are obtained by different involved seismic modeling techniques, whose details  \cite{Pica_A_2005_p-seg_3d_srmmpr} are beyond the scope of the paper. The core of adaptive multiple removal techniques in seismic data boils down to locally  adapt the patterns in Fig. \ref{fig:real2D_signal}-(b) in location and amplitude to the data in Fig. \ref{fig:real2D_signal}-(a). Once adapted, the approximate patterns may be subtracted from the observed signal, with the hope of unveiling previously hidden signals.

\begin{figure}[htb]
\centering
\subfigure[]{\includegraphics[width=8.8cm]{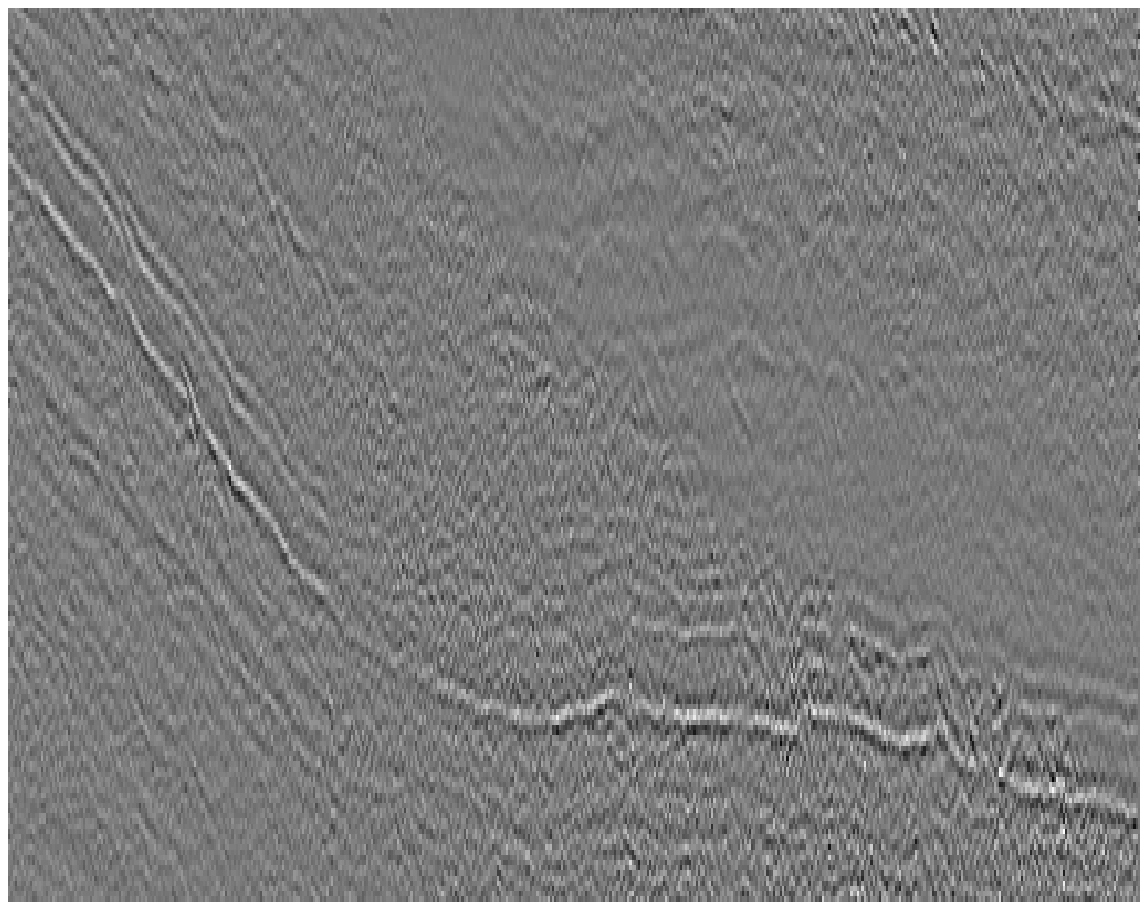}} \hspace{0.05mm}
\subfigure[]{\includegraphics[width=8.8cm]{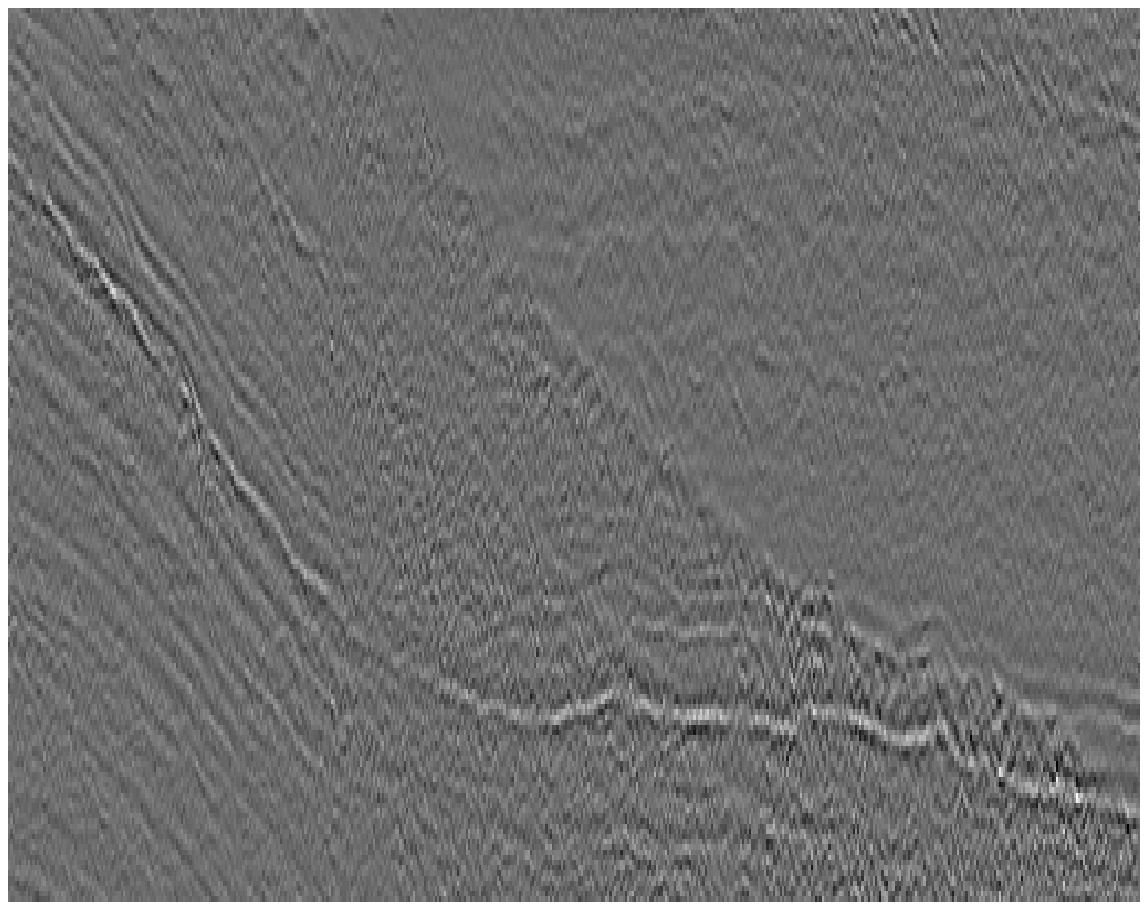}}\\
\caption{Subtraction results, low field-noise case:   primaries (separated from multiples) with (a) \cite{Ventosa_S_2012_j-geophysics_adaptive_mswbcuwf} (b) with the proposed method.  \label{fig:real2D_results_Noise000}} 
\end{figure}

The efficiency of a seismic data processing algorithm is difficult to assess, due to the absence of ground truth. One of the challenges of present seismic data processing resides in the ability to identify deeper target. To this aim, either noisier data sets or broadband seismic acquisitions are being address by  geophysical signal processing.
Fig.~\ref{fig:real2D_results_Noise000} thus compares the results obtained with \cite{Ventosa_S_2012_j-geophysics_adaptive_mswbcuwf} and the proposed algorithm. Although the random noise is apparently highly heteroskedastic, both methods are able to successfully retrieve the weak primary below the multiple level, especially of the left side of the figure.  The method in \cite{Ventosa_S_2012_j-geophysics_adaptive_mswbcuwf} may suffer from a little more pre-echo above the primary in the top-left corner of Fig.~\ref{fig:real2D_results_Noise000}-(a), while a remnant \SI{-45}{\degree} shadow affects the proximal multiple removal in its central part of Fig.~\ref{fig:real2D_results_Noise000}-(b).

\begin{figure}[htb]
\centering
\subfigure[]{\includegraphics[width=8.8cm]{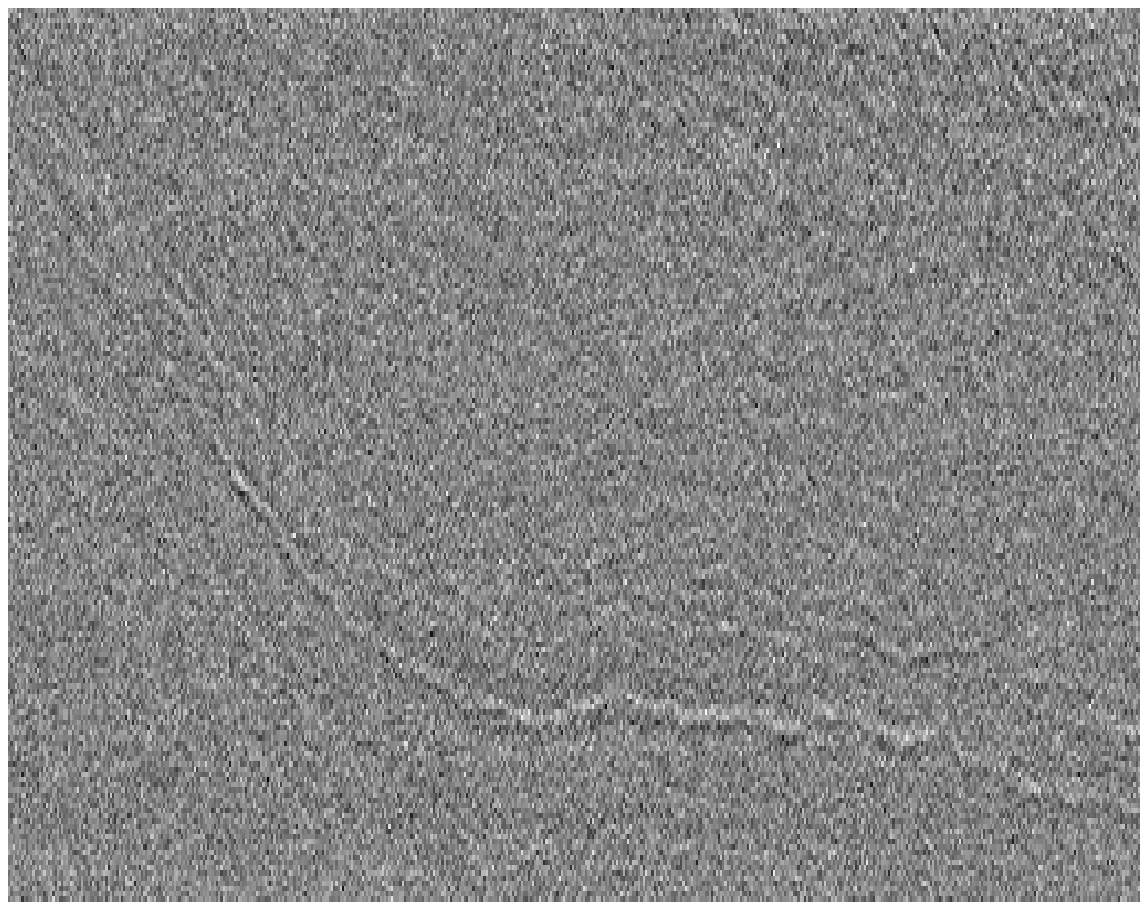}} \hspace{0.05mm}
\subfigure[]{\includegraphics[width=8.8cm]{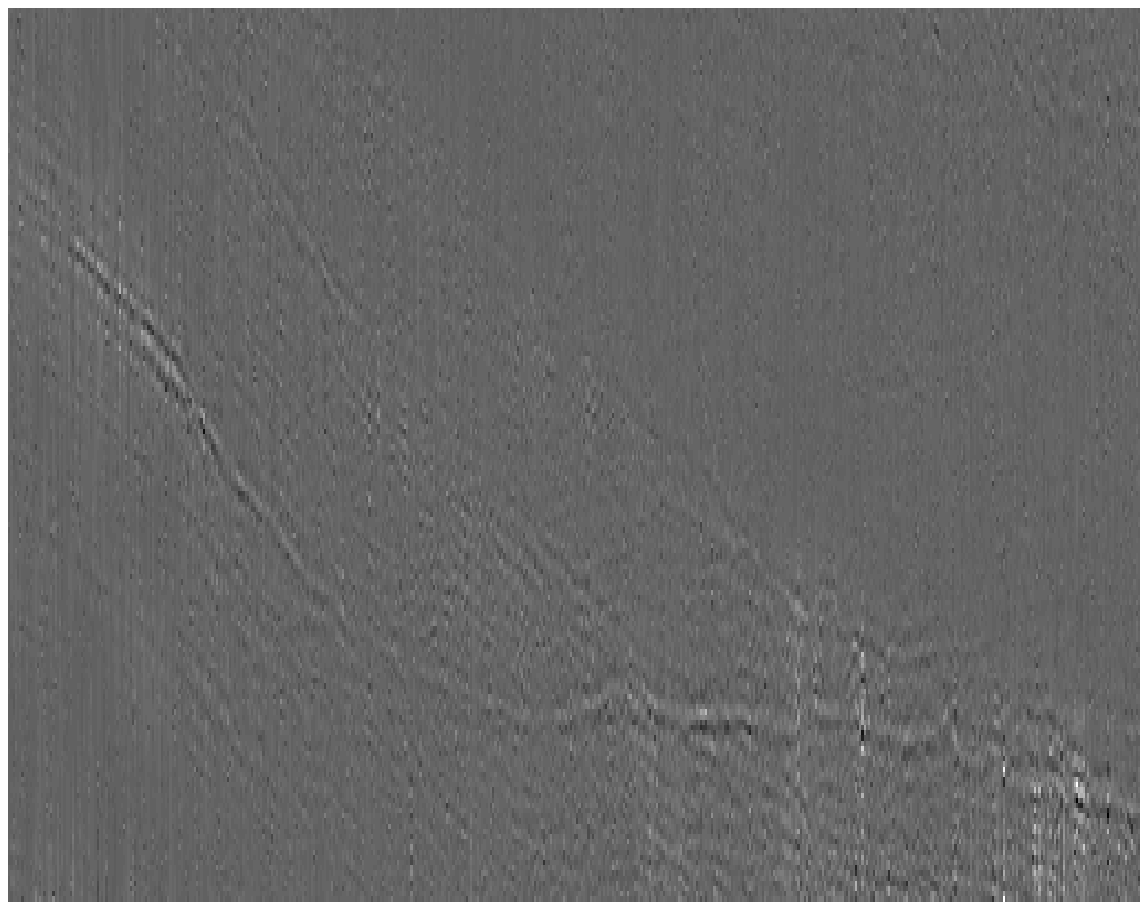}}\\
\caption{Subtraction results, high field-noise case:   primaries (separated from multiples) with (a) \cite{Ventosa_S_2012_j-geophysics_adaptive_mswbcuwf} (b) with the proposed method.  \label{fig:real2D_results_Noise004}} 
\end{figure}

The increased robustness to noisier seismic data is estimated with  a wide-band Gaussian noise,  added to the seismic field data. The outcome is illustrated in Fig.~\ref{fig:real2D_results_Noise004}. While the primary can still be tracked with  \cite{Ventosa_S_2012_j-geophysics_adaptive_mswbcuwf} in Fig.~\ref{fig:real2D_results_Noise004}-(a), it dims inside the ambient noise on the left-most side. The proposed template-based multiple filtering is more robust to noise, reflecting its practical potential. Naturally, the anisotropic, oriented nature of seismic data, and the directional diversity of primaries and multiples, suggests an extension to oriented frames in two dimensions.

\section{Conclusions}
\label{sec:conclusion}
We have proposed a generic methodology to impose sparsity and regularity properties through constrained adaptive filtering in a  transformed domain. This method exploits side information from approximate disturbance templates. The employed proximal framework 
permits different strategies for sparse modeling, additive noise removal, and adaptive filter design under appropriate
regularity and 
amplitude  
coefficient concentration constraints. 
The proposed approach is evaluated on seismic data using different orthogonal wavelet bases and tight frames, and various sparsity measures for wavelet coefficients. 
The standard sparsity-prone $\ell_{1}$-norm is usefully complemented by alternative concentration measures, such as $\ell_{2}$ or $\ell_{1,2}$-norms, which seem better suited to  adaptive filter design.
Its performance is interesting for instance in recovering weak signals buried under both strong random and structured noise. Provided appropriate templates are obtained, this structured-pattern filtering algorithm could be useful in other application areas, e.g. acoustic echo-cancellation in sound and speech,  non-destructive
testing where transmitted waves may rebound at material interfaces (e.g. ultrasounds), or pattern matching in images.
In our future work, two-dimensional directional multiscale approaches \cite{Jacques_L_2011_j-sp_panorama_mgrisdfs} may provide sparser representations for seismic data.
Second, sparsity priors could be enforced on multiple signals as well, with a need for more automation in optimal choices on loss functions in the proximal formulation,  potentially by using other measures than $\ell_1$ \cite{Selesnick_I_2014_j-ieee-tsp_sparse_semsco,Marjanovic_G_2013_p-icassp_exact_lqd,Cetin_A_2013_p-globalsip_projections_ocspocsbol}.  
The Bayesian framework provided in this work could also serve to develop other statistical approaches for multiple removal, e.g.
by using Markov Chain Monte-Carlo methods.

\section*{Acknowledgements}
\label{sec:acknowledgements}
The authors thank Bruno L\'ety and Jean Charl\'ety (IFP Energies nouvelles) for their careful proofreading and result analysis, as well as the reviewers for their detailed suggestions and corrections that improved the manuscript.

\end{document}